\documentclass[symmetry,article,accept,pdftex,moreauthors]{Definitions/mdpi} 
\firstpage{1} 
\pubvolume{1}
\issuenum{1}
\articlenumber{0}
\pubyear{2023}
\copyrightyear{2023}
\externaleditor{{Academic Editor: Xiongfei Wang} %MDPI: please complete this part.
} % For journal Automation, please change Academic Editor to "Communicated by"
\datereceived{30 January 2023} 
\daterevised{13 February 2023}
\dateaccepted{14 February 2023} 
\datepublished{} 
\hreflink{https://doi.org/} % If needed use \linebreak
\doinum{10.3390/sym15020522}
\pdfoutput=1

\usepackage{ifthen}
\newboolean{pdflatex}
\setboolean{pdflatex}{true} 
\newboolean{articletitles}
\setboolean{articletitles}{true} 
\newboolean{uprightparticles}
\setboolean{uprightparticles}{false} 

\usepackage{xspace} 
\usepackage{upgreek}

\def\lhcb {\mbox{LHCb}\xspace}

\def\st {ST\xspace}

\def\MagUp {\mbox{\em Mag\kern -0.05em Up}\xspace}

\ifthenelse{\boolean{uprightparticles}}%
{

	\def\Pmu   {\ensuremath{\upmu}\xspace}

	\def\Ppi   {\ensuremath{\uppi}\xspace}

	\def\Ppsi  {\ensuremath{\uppsi}\xspace}

	\def\PDelta  {\ensuremath{\Delta}\xspace}     
	\def\PXi   {\ensuremath{\Xi}\xspace}     
	\def\PLambda {\ensuremath{\Lambda}\xspace}     
	\def\PSigma  {\ensuremath{\Sigma}\xspace}     
	\def\POmega  {\ensuremath{\Omega}\xspace}     
	\def\PUpsilon {\ensuremath{\Upsilon}\xspace}

	\def\PB  {\ensuremath{\mathrm{B}}\xspace}     
	     
	\def\PD  {\ensuremath{\mathrm{D}}\xspace}

	\def\PJ  {\ensuremath{\mathrm{J}}\xspace}     
	\def\PK  {\ensuremath{\mathrm{K}}\xspace}

	\def\PW  {\ensuremath{\mathrm{W}}\xspace}

	\def\Pb  {\ensuremath{\mathrm{b}}\xspace}     
	\def\Pc  {\ensuremath{\mathrm{c}}\xspace}     
	\def\Pd  {\ensuremath{\mathrm{d}}\xspace}

	\def\Ph  {\ensuremath{\mathrm{h}}\xspace}     
	\def\Pi  {\ensuremath{\mathrm{i}}\xspace}

	\def\Pp  {\ensuremath{\mathrm{p}}\xspace}

	\def\Ps  {\ensuremath{\mathrm{s}}\xspace}     
	     
	\def\Pu  {\ensuremath{\mathrm{u}}\xspace}

	\def\thebaroffset{0.0em}
}
{

	\def\Pmu   {\ensuremath{\mu}\xspace}

	\def\Ppi   {\ensuremath{\pi}\xspace}

	\def\Ppsi  {\ensuremath{\psi}\xspace}     
	     
	\mathchardef\PDelta="7101
	\mathchardef\PXi="7104
	\mathchardef\PLambda="7103
	\mathchardef\PSigma="7106
	\mathchardef\POmega="710A
	\mathchardef\PUpsilon="7107
	     
	\def\PB  {\ensuremath{B}\xspace}     
	     
	\def\PD  {\ensuremath{D}\xspace}

	\def\PJ  {\ensuremath{J}\xspace}     
	\def\PK  {\ensuremath{K}\xspace}

	\def\PW  {\ensuremath{W}\xspace}

	\def\Pb  {\ensuremath{b}\xspace}     
	\def\Pc  {\ensuremath{c}\xspace}     
	\def\Pd  {\ensuremath{d}\xspace}

	\def\Ph  {\ensuremath{h}\xspace}     
	\def\Pi  {\ensuremath{i}\xspace}

	\def\Pp  {\ensuremath{p}\xspace}

	\def\Ps  {\ensuremath{s}\xspace}     
	     
	\def\Pu  {\ensuremath{u}\xspace}

	\def\thebaroffset{0.18em}
}
\newcommand{\offsetoverline}[2][\thebaroffset]{\kern #1\overline{\kern -#1 #2}}%

\makeatletter
\ifcase \@ptsize \relax% 10pt
\newcommand{\miniscule}{\@setfontsize\miniscule{4}{5}}% \tiny: 5/6
\or% 11pt
\newcommand{\miniscule}{\@setfontsize\miniscule{5}{6}}% \tiny: 6/7
\or% 12pt
\newcommand{\miniscule}{\@setfontsize\miniscule{5}{6}}% \tiny: 6/7
\fi
\makeatother

\DeclareRobustCommand{\optbar}[1]{\shortstack{{\miniscule (\rule[.5ex]{1.25em}{.18mm})}
		\\ [-.7ex] $#1$}}

\def\mun  {{\ensuremath{\Pmu^-}}\xspace} % muon negative (\mum is taken)

\def\mumu  {{\ensuremath{\Pmu^+\Pmu^-}}\xspace}

\def\Wpm {{\ensuremath{\PW^\pm}}\xspace}

\def\uquark {{\ensuremath{\Pu}}\xspace}
\def\uquarkbar {{\ensuremath{\overline \uquark}}\xspace}

\def\dquark {{\ensuremath{\Pd}}\xspace}
\def\dquarkbar {{\ensuremath{\overline \dquark}}\xspace}

\def\squark {{\ensuremath{\Ps}}\xspace}
\def\squarkbar {{\ensuremath{\overline \squark}}\xspace}

\def\cquark {{\ensuremath{\Pc}}\xspace}

\def\bquark {{\ensuremath{\Pb}}\xspace}

\def\hadron {{\ensuremath{\Ph}}\xspace}
\def\pion {{\ensuremath{\Ppi}}\xspace}

\def\pip {{\ensuremath{\pion^+}}\xspace}
\def\pim {{\ensuremath{\pion^-}}\xspace}

\def\kaon {{\ensuremath{\PK}}\xspace}
\def\KorKbar {\kern \thebaroffset\optbar{\kern -\thebaroffset \PK}{}\xspace}

\def\Kp  {{\ensuremath{\kaon^+}}\xspace}
\def\Km  {{\ensuremath{\kaon^-}}\xspace}

\def\KS  {{\ensuremath{\kaon^0_{\mathrm{S}}}}\xspace}

\def\Dbar {{\ensuremath{\offsetoverline{\PD}}}\xspace}
\def\D  {{\ensuremath{\PD}}\xspace}

\def\DorDbar {\kern \thebaroffset\optbar{\kern -\thebaroffset \PD}\xspace}
\def\Dz  {{\ensuremath{\D^0}}\xspace}
\def\Dzb {{\ensuremath{\Dbar{}^0}}\xspace}

\def\B  {{\ensuremath{\PB}}\xspace}

\def\BorBbar {\kern \thebaroffset\optbar{\kern -\thebaroffset \PB}\xspace}

\def\Bd  {{\ensuremath{\B^0}}\xspace}

\def\BdorBdbar {\kern \thebaroffset\optbar{\kern -\thebaroffset \Bd}\xspace}
\def\Bu  {{\ensuremath{\B^+}}\xspace}

\def\Bp  {{\ensuremath{\Bu}}\xspace}

\def\Bs  {{\ensuremath{\B^0_\squark}}\xspace}

\def\BsorBsbar {\kern \thebaroffset\optbar{\kern -\thebaroffset \Bs}\xspace}

\def\jpsi {{\ensuremath{{\PJ\mskip -3mu/\mskip -2mu\Ppsi\mskip 2mu}}}\xspace}

\def\Y#1S{\ensuremath{\PUpsilon{(#1S)}}\xspace}

\def\proton  {{\ensuremath{\Pp}}\xspace}
\def\antiproton {{\ensuremath{\overline \proton}}\xspace}

\def\Lz   {{\ensuremath{\PLambda}}\xspace}
\def\Lbar  {{\ensuremath{\offsetoverline{\PLambda}}}\xspace}
\def\LorLbar {\kern \thebaroffset\optbar{\kern -\thebaroffset \PLambda}\xspace}

\def\Sigmares {{\ensuremath{\PSigma}}\xspace}

\def\Xires  {{\ensuremath{\PXi}}\xspace}

\def\Xiresbar {{\ensuremath{\offsetoverline{\Xires}}}\xspace}

\def\Lc   {{\ensuremath{\Lz^+_\cquark}}\xspace}

\def\Xicp  {{\ensuremath{\Xires^+_\cquark}}\xspace}

\def\Lb   {{\ensuremath{\Lz^0_\bquark}}\xspace}
\def\Lbbar  {{\ensuremath{\Lbar{}^0_\bquark}}\xspace}

\def\Xibz   {{\ensuremath{\Xires^0_\bquark}}\xspace}

\def\Xibm   {{\ensuremath{\Xires^-_\bquark}}\xspace}

\def\Xibbarp  {{\ensuremath{\Xiresbar{}_\bquark^+}}\xspace}

\def\Hb {{\ensuremath{H_\bquark}}\xspace}
\def\Hbbar {{\ensuremath{\offsetoverline{H}_\bquark}}\xspace}

\def\to     {\ensuremath{\rightarrow}\xspace}

\def\CP     {{\ensuremath{C\!P}}\xspace}

\def\CPV     {{\ensuremath{C\!PV}}\xspace}

\def\Vub {{\ensuremath{V_{\uquark\bquark}}}\xspace}
\def\Vcb {{\ensuremath{V_{\cquark\bquark}}}\xspace}

\def\Vuss {{\ensuremath{V_{\uquark\squark}^\ast}}\xspace}
\def\Vcss {{\ensuremath{V_{\cquark\squark}^\ast}}\xspace}

\def\dacp		{\ensuremath{\Delta A_{CP}}\xspace}

\newcommand{\AT}{{\ensuremath{A_{\widehat{T}}}}\xspace}
\newcommand{\ATbar}{{\ensuremath{\kern 0.1em\overline{\kern -0.1em A}_{\widehat{T}}}}\xspace}
\newcommand{\aCPTodd}{{\ensuremath{a_\CP^{\widehat{T}\text{-odd}}}}\xspace}
\newcommand{\aPTodd}{{\ensuremath{a_P^{\widehat{T}\text{-odd}}}}\xspace}

\newcommand{\CT}{{\ensuremath{C_{\widehat{T}}}}\xspace}
\newcommand{\CTbar}{{\ensuremath{\kern 0.1em\overline{\kern -0.1em C}_{\widehat{T}}}}\xspace}

\def\That  {\ensuremath{\widehat{T}}\xspace}

\def\C#1  {\ensuremath{\mathcal{C}_{#1}}\xspace}       % 9
\def\Cp#1 {\ensuremath{\mathcal{C}_{#1}^{'}}\xspace}      % 7
\def\Ceff#1 {\ensuremath{\mathcal{C}_{#1}^{\mathrm{(eff)}}}\xspace}  % 9 
\def\Cpeff#1 {\ensuremath{\mathcal{C}_{#1}^{'\mathrm{(eff)}}}\xspace}  % 7
\def\Ope#1 {\ensuremath{\mathcal{O}_{#1}}\xspace}       % 2
\def\Opep#1 {\ensuremath{\mathcal{O}_{#1}^{'}}\xspace}      % 7

\newcommand{\aunit}[1]{\ensuremath{\text{\,#1}}}  
\newcommand{\tev}{\aunit{Te\kern -0.1em V}\xspace}
\newcommand{\gev}{\aunit{Ge\kern -0.1em V}\xspace}
\newcommand{\mev}{\aunit{Me\kern -0.1em V}\xspace}
\newcommand{\kev}{\aunit{ke\kern -0.1em V}\xspace}
\newcommand{\ev}{\aunit{e\kern -0.1em V}\xspace}
\newcommand{\mevc}{\ensuremath{\aunit{Me\kern -0.1em V\!/}c}\xspace}
\newcommand{\gevc}{\ensuremath{\aunit{Ge\kern -0.1em V\!/}c}\xspace}
\newcommand{\mevcc}{\ensuremath{\aunit{Me\kern -0.1em V\!/}c^2}\xspace}
\newcommand{\gevcc}{\ensuremath{\aunit{Ge\kern -0.1em V\!/}c^2}\xspace}
\newcommand{\gevgevcccc}{\ensuremath{\gev^2\!/c^4}\xspace} % for q^2

\def\fb {\ensuremath{\aunit{fb}}\xspace}
\def\invfb {\ensuremath{\fb^{-1}}\xspace}

\def\gsim{{~\raise.15em\hbox{$>$}\kern-.85em
		\lower.35em\hbox{$\sim$}~}\xspace}
\def\lsim{{~\raise.15em\hbox{$<$}\kern-.85em
		\lower.35em\hbox{$\sim$}~}\xspace}

\def\sqs {\ensuremath{\protect\sqrt{s}}\xspace}

\def\pt   {\ensuremath{p_{\mathrm{T}}}\xspace}

\def\tell1 {TELL1\xspace}
\def\ukl1 {UKL1\xspace}

\Title{$CP$ Violation in Baryon Decays at LHCb}

\TitleCitation{$CP$ Violation in Baryon Decays at LHCb}

\Author{Xinchen Dai %Please carefully check the accuracy of names and affiliations. %Re: the names and affiliations are correct
*\orcidA{}, Miroslav Saur *\orcidM{}, Yiduo Shang \orcidY{}, Xueting Yang \orcidX{} and Yanxi Zhang *\orcidZ{}}

\AuthorNames{Xinchen Dai, Miroslav Saur, Yiduo Shang, Xueting Yang and Yanxi Zhang}

\AuthorCitation{Dai, X.; Saur, M.; Shang, Y.; Yang, X.; Zhang, Y.}
\address[1]{%
School of Physics, Peking University, Beijing 100871, China; shangyiduo@stu.pku.edu.cn (Y.S.); yangxt@stu.pku.edu.cn (X.Y.)}

\corres{\hangafter=1 \hangindent=1.05em \hspace{-0.82em}Correspondence: xinchen.dai@pku.edu.cn (X.D.); saurm@pku.edu.cn (M.S.); yanxi.zhang@pku.edu.cn (Y.Z.) }

\abstract{
Observations in astronomy suggest that our Universe contains an abundance of matter over antimatter, which can only be explained if the combined \CP symmetry is violated. Studies of \CP violation have driven the flavor physics with the aim of testing the Standard Model of particle physics and searching for physics beyond it. \CP violation is discovered in strange, beauty, and charm meson systems; however, no conclusive sign of \CP violation in baryon decays has been observed yet. This review summarizes \CP violation studies performed by the \lhcb experiment in charmless decays and rare decays of beauty baryons and singly Cabibbo-suppressed decays of charm baryons. 
A brief on prospects for future \lhcb measurements is also discussed.
}


\begin{document}
\section{Introduction}
\label{sec:introduction}
As stated in the Noether's Theorem, the conservation of a quantity corresponds to a continuous symmetry in the underlying interaction~\cite{Noether:1918zz}. Studies of symmetries have played an important role in establishing our understandings of the fundamental dynamics and matter of Nature. 
Besides continuous symmetries, Nature also exhibits discrete symmetries, for example the parity (\textit{P}) and charge conjugation (\textit{C}) symmetries.
Conserving the \textit{P} symmetry, a system is invariant to the space inversion. Similarly, the \textit{C} symmetry links together the properties of particles and their relevant anti-particles. These symmetries are not preserved in all known fundamental interactions, as first postulated for the \textit{P} symmetry by T. D. Lee and C. N. Yang~\cite{Lee:1956qn} and subsequently observed by C. S. Wu in 1956~\cite{Wu:1957my} in weak interactions.
The combined \CP symmetry was unexpectedly observed to be violated in decays of neutral kaons in 1964~\cite{Cronin:1964fg}.
These observations were incorporated into the electroweak theory and subsequently into the Standard Model (SM) of particle physics, via so-called Cabibbo--Kobayashi--Maskawa (CKM)~mechanism~\cite{Kobayashi:1973fv}.

The \CP violation (\CPV) in the SM arises from the \Wpm boson-mediated charged-current interaction with quarks. The quark eigenstates interacting with the \Wpm bosons do not align with those interacting with the Higgs boson. The $3\times3$ unitary CKM matrix:
\begin{equation}
V_{\text{CKM}}\equiv
\begin{pmatrix}
V_{ud} & V_{us} & V_{ub} \\
V_{cd} & V_{cs} & V_{cb} \\
V_{td} & V_{ts} & V_{tb} 
\end{pmatrix},
\end{equation}
proposed by M. Kobayashi and T. Maskawa, describes the mixing between the two eigenstates for three generations of quarks in the SM. Being complex, the CKM matrix has an irreducible weak phase, whose sign is flipped for antiparticles, providing the only established source of \CPV in Nature.

The violation of the \CP symmetry is required to explain the observed asymmetry between baryons and antibaryons in the Universe, often called the Baryon Asymmetry of the Universe (BAU)~\cite{Sakharov:1967dj}. Even though the SM allows \CPV, the CKM mechanism is not strong enough to account for the BAU, demanding new physics for additional sources of \CPV. The last two decades have witnessed great achievements of \CP studies. The two B-factories, BaBar~\cite{BaBar:2001yhh} and Belle~\cite{Belle:2000cnh}, successfully established the CKM mechanism through measurements in $B$ meson decays. The LHCb experiment~\cite{LHCb_detector} extended the measurements to $B_s^0$ and beauty baryons' decays and manifestly increased the precision of measurements for charm hadrons. All obtained results in general agree with predictions in the SM as suggested by a global fit of the CKM parameters~\cite{HFLAV:2022pwe}; however, there is still room for new physics effects. Besides, up to now, \CPV has not been detected in baryon decays despite being driven by dynamics identical to mesons in the SM. \CPV in the baryon sector remains a largely unexplored area.

In this article, measurements of \CPV in beauty and charm baryons at the \lhcb were reviewed. The remaining parts are organized as follows: in Section~\ref{sec:lhcb}, the LHCb experiment is briefly discussed; 
in Section~\ref{sec:exp_methods}, common methods for \CPV measurements are summarized; 
Sections~\ref{sec:beauty_decays} and~\ref{sec:charm_decays} present the results of \CPV in beauty and charm baryons, respectively, and finally, future measurements are prospected in Section~\ref{sec:prospects}.
{For the numeric quoted results in this review, the first uncertainties are statistical, the second are systematic, and the third, if shown, are due to external inputs. All figures reproduced in this review used to demonstrate the \lhcb detector or measurements were published as open access, allowing their reuse, and the detailed information can be found in the original \lhcb~publications.}

\section{The LHCb Experiment}
\label{sec:lhcb}
\begin{figure}[H]
% \centering
 \includegraphics[width = 0.9\textwidth]{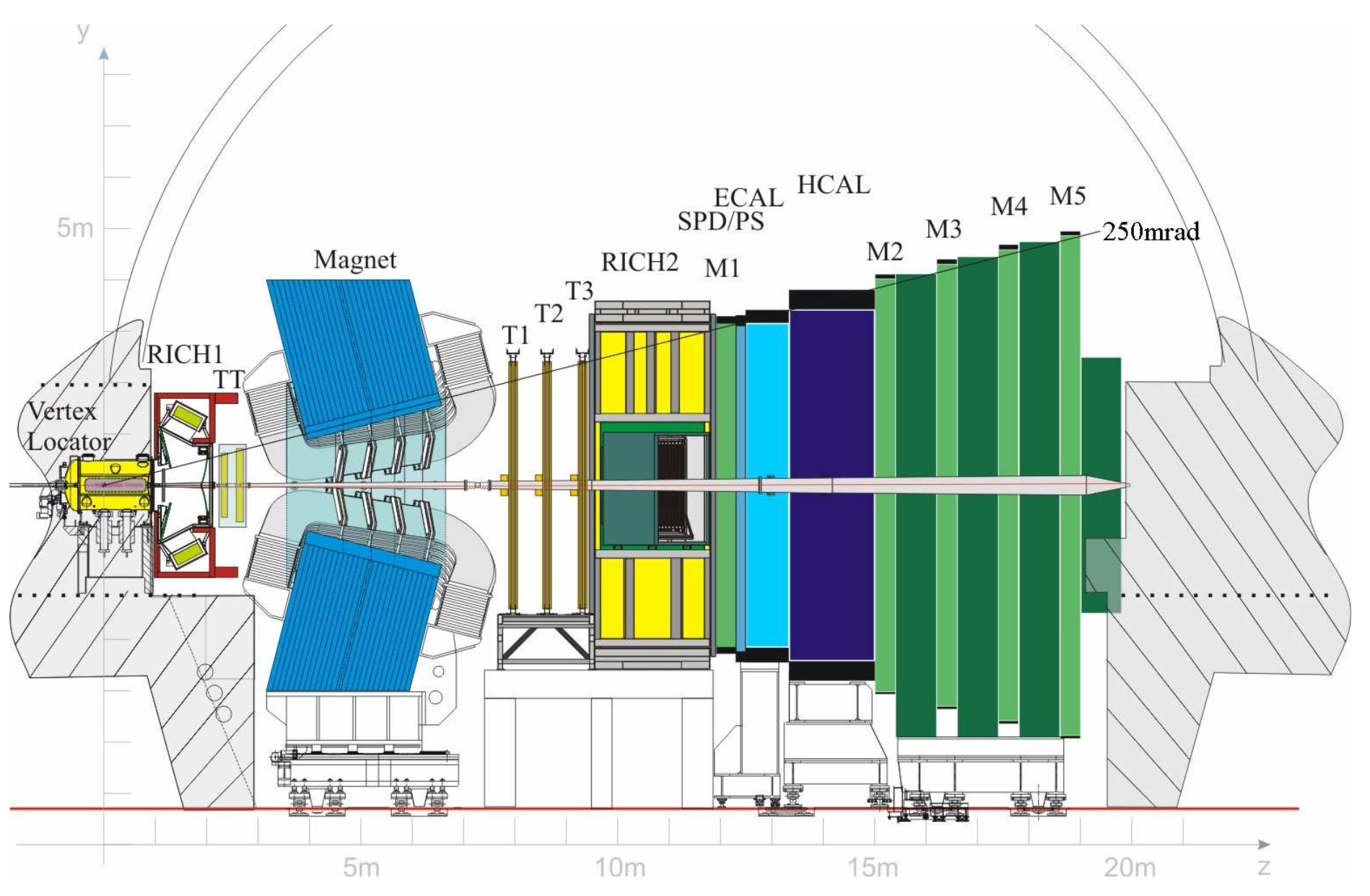}
 \caption{View of the LHCb detector. Reproduced from~\cite{LHCb_perf_Run1}. }
 \label{fig:detector}
\end{figure}

The \lhcb is one of the four main experiments at the Large Hadron Collider (LHC) operated by CERN. It is designed with a focus on precise tracking, vertexing, momentum measurement, and high-efficiency particle identification. It records particles produced in $pp$ collisions within the forward pseudorapidity ($\eta$) range $2<\eta<5$, taking advantage of $b$ and $\overline{b}$ being produced as correlated at relatively small angles with respect to the proton beam direction.
The layout of the LHCb spectrometer during the LHC Run 2 operation is shown in Figure~\ref{fig:detector} and briefly described below. A full description is available in~\cite{LHCb_detector, LHCb_perf_Run1, LHCb_perf_and_trigger_Run2}.
\vspace{-9pt} 

The LHCb magnet provides an integrated magnetic field of about 4 Tm. It deflects charged particles on the horizontal plane to accurately measure their momenta. The magnet field polarity is periodically switched to achieve a comparable amount of data recorded with each polarity, allowing the validation and correction of the instrumental asymmetries of charged particles.
The tracking system consists of the VErtex LOcator (VELO), the Tracker Turicensis (TT), and Tracking Stations T1--T3.
Charged hadron identification is achieved by two Ring Imaging Cherenkov detectors, RICH1 and RICH2.
The calorimeter system consists of a Scintillating Pad Detector, a Preshower Detector, a shashlik-type Electromagnetic Calorimeter (ECAL), and a Hadronic Calorimeter (HCAL). 
The muon detection system is composed of five stations, M1--M5, built predominantly with multiwire proportional chambers, providing muon identification. 

The LHCb trigger system used to collect the data for results reviewed in this article consists of hardware and software stages. The hardware stage combines information from the ECAL and the HCAL to select hadrons with high transverse energy and uses information from the muon system to select muons with high transverse momenta (\pt). 
Selected events are then passed to the software trigger, which utilizes all LHCb systems to perform beauty and charm selections based on information including the momentum, vertex displacement, particle identification, etc.

The LHCb experiment began collecting $pp$ collision data for physics analyses at the start of the LHC's operation in 2010. During the Run 1 period (2010--2012), $pp$ collisions were recorded at center-of-mass energies of $\sqs=7$ or $8\tev$, with an integrated luminosity of $\int\mathcal{L}\approx3\invfb$. For Run 2 (2015--2018), $pp$ collisions were recorded at $\sqs=13\tev$ with $\int\mathcal{L}\approx6\invfb$. The results reviewed in this article were based on these data. 
The Run 3 operation started in 2022 and is expected to finish in 2025. During Run 3, the \lhcb is being operated with a nominal instantaneous luminosity five-times higher than during Run 2.
In order to increase the sensitivity and broaden the accessible physics reach, the LHCb experiment has been significantly upgraded for Run~3~\cite{LHCb_upgrade_TDR}.
This upgrade included a completely new tracking system consisting of a pixel-based VELO and two new tracking detectors, UT and SciFi, which replaced the TT and Tracking Stations T1--T3, respectively.
The trigger system has been upgraded to a fully software one, which allows even higher flexibility in the physics program. 
Hadronic decays, in particular, benefit from this improvement~\cite{LHCb_trigger_TDR}. In total, about $25\invfb$ $pp$ collisions are expected to be collected during Run 3.

Concerning the physics performances, 
the LHCb is specialized in the detection of decays with charged final states, while B-factories are better at decays involving neutral particles. Besides, the measurements of beauty baryons are unique to the LHCb. The next generation of B-factories, the Belle II experiment~\cite{Belle-II:2010dht}, started to accumulate data in 2019. In the next few years, the LHCb and Belle II, complementary to each other, will be the two major experiments with \CP measurements as their core physics program.

\section{Experimental Methods}
\label{sec:exp_methods}
Baryon number conservation prevents mixing of baryons and antibaryons, such that only direct \CPV is allowed in baryon decays, usually quantified as the relative difference between the partial width ($\Gamma$) of a baryon decay $\Gamma(\Hb \to f)$ and its antibaryon decay $\Gamma(\Hbbar \to \overline{f})$:
\begin{equation}
A_\CP^f \equiv \frac{\Gamma(\Hb \to f) - \Gamma(\Hbbar \to \overline{f})}{\Gamma(\Hb \to f) + \Gamma(\Hbbar \to \overline{f})}. \label{eq:a_cp_gamma}
\end{equation}
For \CPV to emerge, %MDPI: please check if the noindent for this paragraph is necessary, if no, please remove it. %Re: there is noindent here
 at least two sub-processes are required to contribute with both different weak and strong phases.
Experimentally, the relative partial width difference is transformed to the relative signal yield difference as
\begin{equation}
A_\CP^\text{raw} \equiv \frac{N(\Hb \to f) - N(\Hbbar \to \overline{f})}{N(\Hb \to f) + N(\Hbbar \to \overline{f})},
\label{eq:a_cp_exp}
\end{equation}
where \ensuremath{N} represents the yield of a specific decay. {Besides the \CP asymmetry $A_\CP$ in $b$-hadron decays, $A_\CP^\text{raw}$ also contains experimental asymmetries, which must be subtracted. Experimental asymmetries include}
the asymmetry of the $\Hb$ and $\Hbbar$ production cross-section {in proton--proton collisions} {(the production asymmetry)} and the asymmetry of the final state $f$ detection efficiency {(the detection asymmetry)}. {These experimental asymmetries would create a false \CP signature if not properly corrected.}
At the LHCb, the production asymmetry of the most-studied $\Lb$ baryon is $1\sim2 \%$, evidencing a trend increasing with the $\Lb$ rapidity~\cite{LHCb:2021xyh}. The detection asymmetry is about $1\%$ for protons~\cite{LHCb:2021xyh} and kaons~\cite{LHCb:2014kcb} and is reduced with a higher momentum. The detection asymmetry for pions is approximately zero~\cite{LHCb:2012swq}. For measurements aiming at a precision below a percent level, it is often useful to measure the difference of the \CP asymmetry between the signal mode and a control mode, \dacp, in order to cancel the experimental asymmetries. The control mode is usually chosen to be one with negligible \CPV.
If the expected \CP asymmetries in relevant channels have a different sign, \dacp also helps to enhance the experimental sensitivity.

For two-body decays, i.e., the final state $f$ is composed of two particles, Equation~\eqref{eq:a_cp_exp} gives the only result that can be probed experimentally, named the global asymmetry. For multi-body decays, $A_\CP$ can be dependent on the phase-space of the final state $f$, such that localized asymmetries measured in different phase-space regions provide additional information to understand the dynamics of \CPV.
There are a few methods to study phase-space-dependent \CPV, varying in the experimental complexity, the sensitivity to \CPV, and the extent of information probed~\cite{Bediaga:2020qxg}.

Measuring \CPV by binning phase-space is a very often-used method, for example that implemented with the Miranda technique~\cite{Bediaga:2012tm}.
Various phase-space binning schemes can be defined, such as uniform binning, where each bin has the same area, adaptive binning, where each bin is required to have the same amount of signal events, and physics-motivated binning, following specific resonance structures of the studied decay channel.
The localized \CPV in each bin then can be measured according to Equation~\eqref{eq:a_cp_exp} as for the global symmetry, which is nicely illustrated by the $\Bp\to h_1h_2h_3$ analysis~\cite{LHCb:2022nyw}. Specifically for the Miranda technique, the asymmetry significance is calculated for the number of baryon ($n_i$) and antibaryon ($\overline{n}_i$) decays for each bin $i$:
\begin{equation}
 S^i_\CP=\frac{n_i-\alpha \overline{n}_i}{\sqrt{\alpha (n_i+ \overline{n}_i)}},
\end{equation}
where $i$ runs over the phase-space bins and $\alpha=\sum_i n_i/\sum_i \overline{n}_i$ normalizes the total number of baryon decays to that of antibaryons and removes global asymmetries. In the case of no \CPV, $S_\CP$ follows a normal distribution.

Unbinned methods, which overcome the ambiguity of binning scheme choices, are also implemented to study \CPV in multi-body decays.
In the so-called Energy Test~\cite{ET1} method, the unbinned model-independent test statistic $T$ measures the average distance between pairs of $\Hb$ and $\Hbbar$ decays, \textit{i} and \textit{j}, based on the selected metric $\psi_{ij}$:
\begin{equation}
T \equiv \frac{1}{2n(n-1)}\sum_{i\neq j}^{n}\psi_{ij}
+ \frac{1}{2\overline{n}(\overline{n}-1)}\sum_{i\neq j}^{\overline{n}}\psi_{ij}
- \frac{1}{n\overline{n}} \sum_{i=1}^{n}\sum_{j=1}^{\overline{n}}\psi_{ij} ,
\label{eqn:energy_test_main}
\end{equation}
where \textit{n} and $\overline{\textit{n}}$ are the signal yields in the $\Hb$ and $\Hbbar$ samples, respectively. The first (second) term sums over the metric-weighted distance of pairs from the $\Hb$ ($\Hbbar$) sample, whereas the third term sums over pairs of an $\Hb$ and an $\Hbbar$ decay. A usual choice for the metric function is $\psi_{ij} = e^{-d^2_{ij}/2\delta^2}$, where $d_{ij}$ is the distance of events \textit{i} and \textit{j} calculated using phase-space variables and $\delta$ is a tunable effective radius defining the scale of the phase-space region of the $\Hb$ and $\Hbbar$ being compared. 
Without \CP asymmetries, \textit{T} would be consistent with 0, while a large \CPV would result in a large positive $T$.
The expected distribution of $\textit{T}$ under null \CPV is usually obtained using permutation samples, in which the flavor of each decay as $\Hb$ or $\Hbbar$ is randomly reassigned. 

The $k$-Nearest Neighbors (kNN) technique~\cite{kNN1,Williams:2010vh} is another unbinned method based on a set of nearest neighbor candidates ($n_k$) in the combined sample of $\Hb$ and $\Hbbar$ decays. The test statistic $T$ is defined as 
\begin{equation}
T \equiv \frac{1}{n_k(n+\overline{n})}\sum_{i=1}^{n+\overline{n}} \sum_{k=1}^{n_k}I(i,k),
\label{eqn:kNN}
\end{equation}
where $I(i,k)=1$ if the $i^{th}$ candidate and its $k^{th}$ nearest neighbor have the same change and $I(i,k)=0$ otherwise. To determine the $k^{th}$ nearest neighbors, a distance metric must be defined as for the Energy Test method. The parameter $n_k$ is tunable to study its sensitivity on the \CPV measurement. The distribution of $T$ under the hypothesis of no \CPV follows a normal distribution with its mean and variance calculated from known parameters.

The amplitude analysis is a powerful unbinned method to search for \CP violation over the phase-space of a multi-body decay~\cite{Back:2017zqt}. The total amplitude $\mathcal{A}$ ($\mathcal{\overline{A}}$) for the $\Hb$ ($\Hbbar$) decay can be described model dependently by the coherent sum of quasi-two-body amplitudes of intermediate resonant or non-resonant contributions:
\begin{equation}
 \mathcal{\overset{(-)}{A}}=\sum_{i}\overset{(-)}{a_i} \overset{(-)}{\mathcal{A}_i}
 \vspace{-6pt} 
\end{equation}
where $\overset{(-)}{\mathcal{A}_i}$ is the amplitude for the intermediate state $i$ and $\overset{(-)}{a_i}$ the corresponding complex coefficient. The amplitude $\mathcal{A}_i$ as a function of the $n$-body phase-space variables $\Phi_n$ can be parameterized by the helicity formalism~\cite{Jacob:1959at} or the covariant tensor formalism~\cite{Zemach:1965ycj}. The probability density function is given by~\cite{LHCb:2019sus}
\begin{equation}
 \mathcal{S}(\Phi_n;a_i, \overline{a}_i)=\frac{\epsilon[ (1+q)|\mathcal{A}|^2+(1-q)|\mathcal{\overline{A}}|^2]}{\int \epsilon [(1+q)|\mathcal{A}|^2+(1-q)|\mathcal{\overline{A}}|^2]d\Phi_n},\label{eq:amp_model}
\end{equation}
where $\epsilon$ represents the experimental efficiency and $q$ equals $+1$ and $-1$ for $\Hb$ and $\Hbbar$, respectively. The coefficients $a_i, \overline{a}_i$ of each component are determined by fitting to the data, and the differences in the magnitudes and phases of $a_i$ and $\overline{a}_i$ represent \CPV. The relative \CPV can also be defined as
\begin{equation}
 \mathcal{A}^i_{\CP}= \frac{|a_i|^2-|\overline{a}_i|^2}{|a_i|^2+|\overline{a}_i|^2}.\label{eq:CPV_amp}
\end{equation}
There are a few known complexities associated with amplitude analyses, for example the compositions of resonances in the amplitude, the large consumption of computing resources, and possible multiple solutions. A model-independent moment-like analysis was proposed in~\cite{Zhang:2022emj} to deal with these difficulties. Data are projected onto a series of orthogonal Legendre polynomials of an angular phase-space variable, and any differences between the coefficients of the $\Hb$ and $\Hbbar$ polynomials signals \CPV. %AE: please check if there need an indent. %Re: We don't need an indent here

For the four-body decays $\Hb\to ph_1h_2h_3, h=K,\pi$, the Triple Product Asymmetry (TPA)~\cite{TPA_Bensalem:2002ys,TPA_Gronau:2015gha,Geng:2021sxe,Wang:2022fih} is also used to measure the \CPV at \lhcb. The triple product for an $\Hb$ (\CT) or $\Hbbar$ (\CTbar) decay is calculated using final state momenta, $\vec{p}_i$, evaluated in the $b$-baryon rest frame as
\begin{align}
\CT\equiv\vec{p}_\proton\cdot(\vec{p}_{h_1}\times\vec{p}_{h_2}),~~
\CTbar\equiv\vec{p}_\antiproton\cdot(\vec{p}_{\overline{h}_1}\times\vec{p}_{\overline{h}_2}),
\end{align}
where $\That$ is an operator that flips the direction of final state momenta and spin.
Then the TPA is calculated for both $\Hb$ and $\Hbbar$ decays
\begin{align}
\label{eq:tpa_at_ct}
\AT(\CT) &= \frac{N(\CT>0)-N(\CT<0)}{N(\CT>0)+N(\CT<0)},\\
%\quad \quad \quad
\label{eq:tpa_at_ctbar}
\ATbar(\CTbar) &= \frac{\overline{N}(-\CTbar>0)-\overline{N}(-\CTbar<0)}{\overline{N}(-\CTbar>0)+\overline{N}(-\CTbar<0)},
\end{align}
where $N$ and $\overline{N}$ are the yields of \Hb and \Hbbar decays that have the triple product with a definite sign.
The TPA is $P$-odd and $\That$-odd, namely it changes sign under $P$ or \That transformations, \mbox{$\AT\overset{P,\That}{=} - \AT$}.
From the TPA, the following $P$- and \CP-violating observables are defined:
\begin{align}
\aPTodd = \frac{1}{2}\left(\AT +\ATbar\right),\\
\aCPTodd = \frac{1}{2}\left(\AT- \ATbar\right),
\label{eq:tpa_atp_atcp}
\end{align}
The advantage of measuring \CPV using the triple product asymmetry is that it is insensitive to instrumental asymmetries and is complementary to the usual asymmetries due to a different dependence on strong phases. %AE: please check if there need an indent. %Re: We don't need an indent here

%%%%%%%%%%%
\section{Results on Beauty Baryon \CPV}
\label{sec:beauty_decays}
\subsection{\CPV in Charmless Two-Body Decays}
\label{sec:two_body_decays}
Charmless decays of $b$-hadrons are governed by \mbox{ $b\to u(\uquarkbar d/s)$ } and \mbox{ $b\to d/s (q\overline{q})$} diagrams, where \mbox{ $q\overline{q}=u\uquarkbar,d\dquarkbar,s\squarkbar$}. The two sub-processes have different weak and strong phases, so a potential \CPV can arise. The charmless two-body decays $\Lb\to p \Km,p\pi^-$ are in analogy to $B\to h_1h_2$ decays, for which \CPV as large as $\sim$10\% has been observed~\cite{PDG}.
The \CP violation defined according to Equation~\eqref{eq:a_cp_gamma} as $A_\CP[\Lb\to p \Km(\pi^-)]$ is predicted as $5.8\%$ ($-3.9\%$) in the generalized factorization approach~\cite{Hsiao:2014mua} and is $-5\%$ ($-31\%$) in the perturbative QCD approach~\cite{Lu:2009cm}. 
Predictions of large \CPV are inconsistent with the measurements by the LHCb, giving~\cite{LHCb-PAPER-2018-025}
\begin{equation}
\begin{aligned}
& A_{\CP}(\Lambda^0_b \to p\Km)=-0.020\pm0.013\pm0.019, \\
& A_{\CP}(\Lambda^0_b \to p\pi^-)=-0.035\pm0.017\pm0.020.
\end{aligned}
\end{equation}

\begin{figure}[H]
% \centering
 \includegraphics[width=0.36\textwidth]{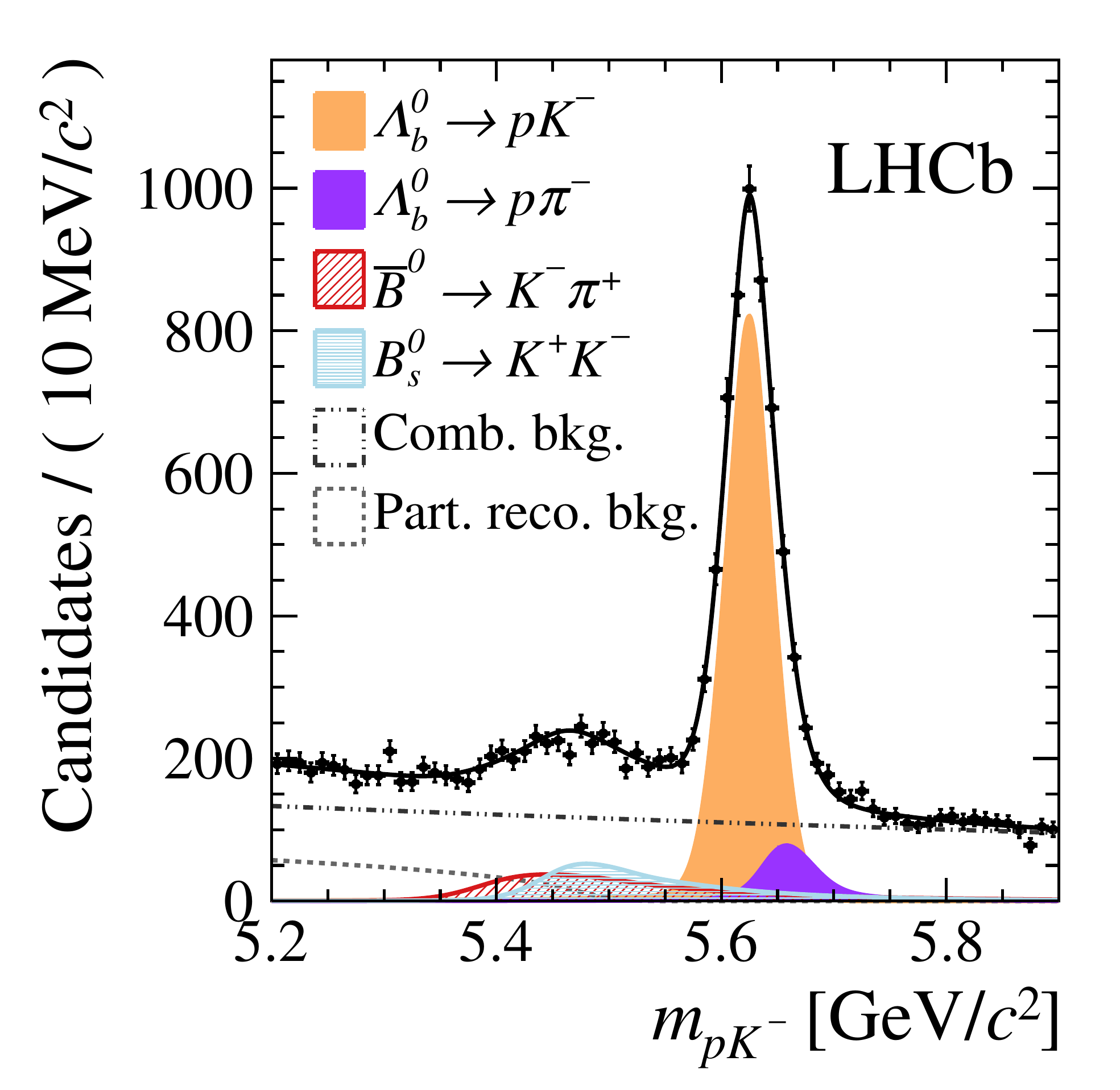}
 \includegraphics[width=0.36\textwidth]{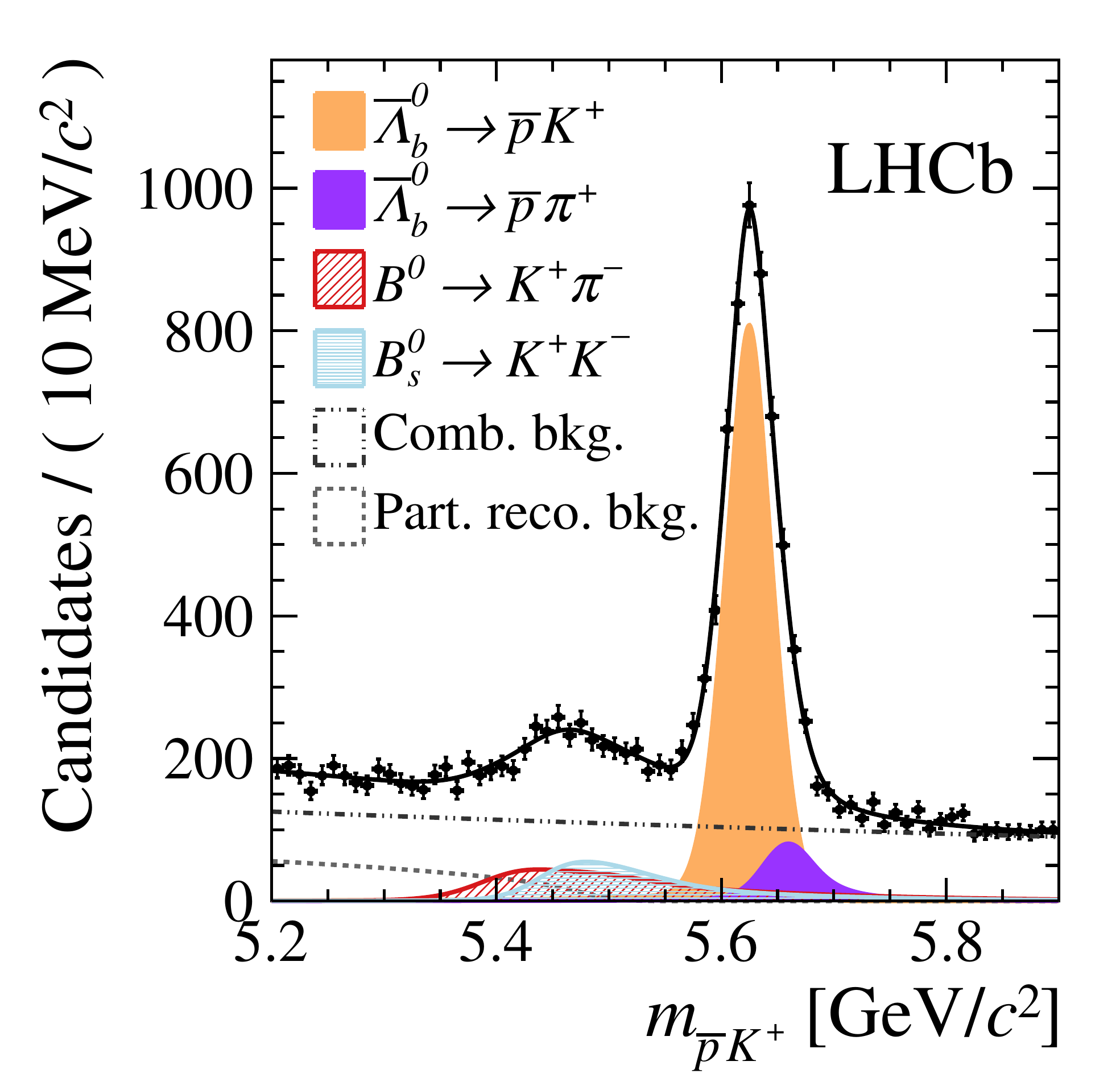}\\
 \includegraphics[width=0.36\textwidth]{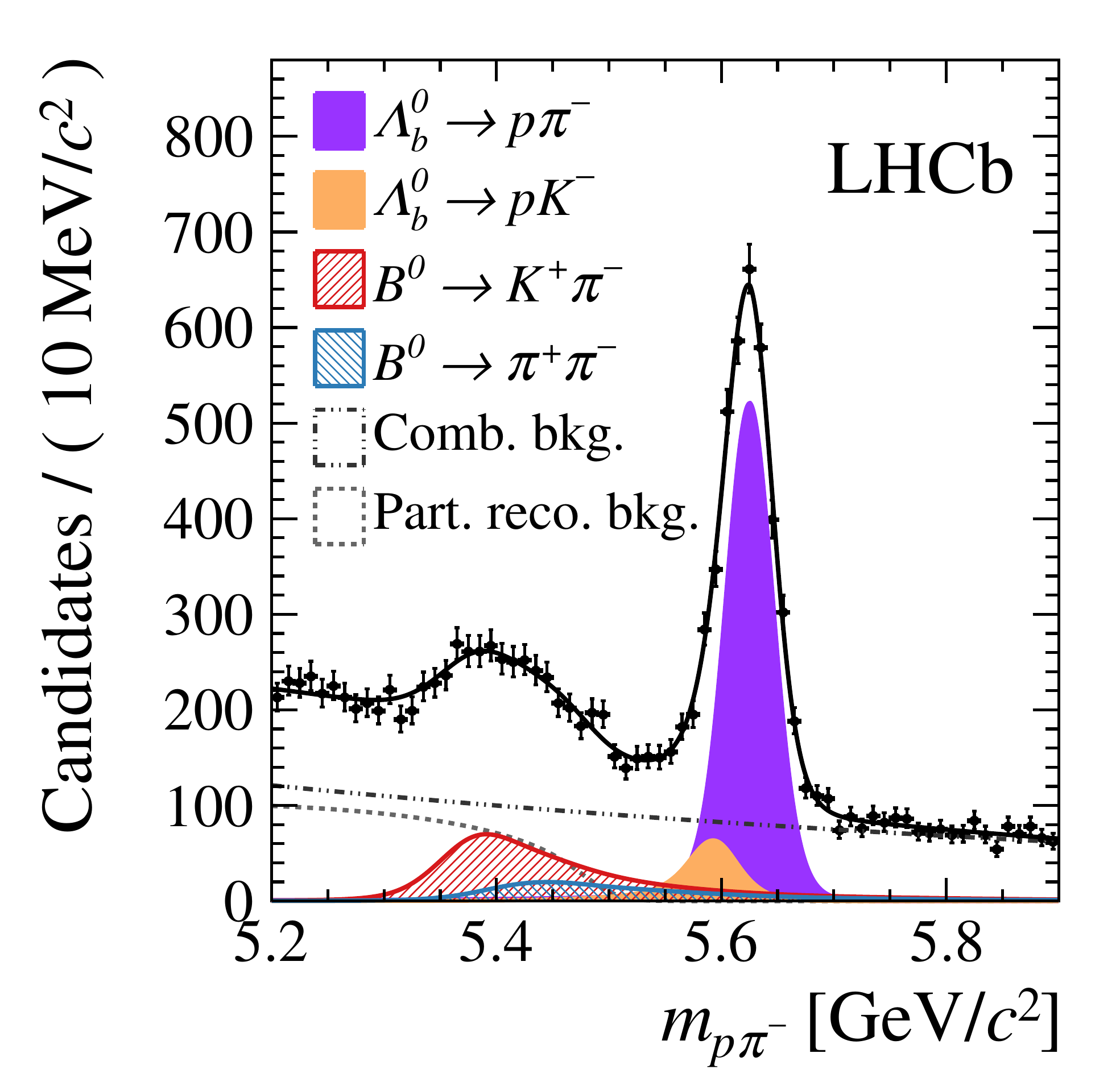}
 \includegraphics[width=0.36\textwidth]{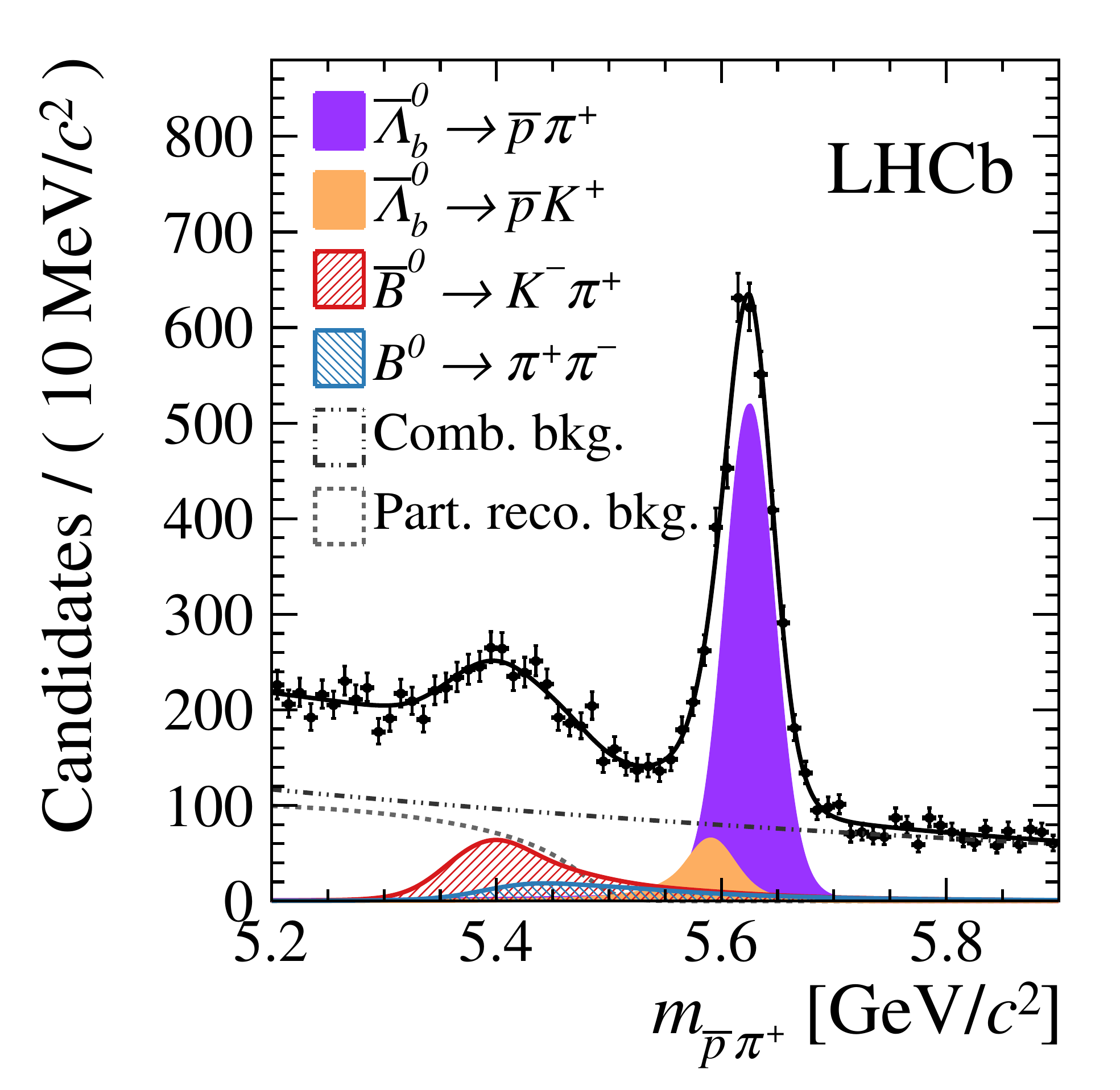}
 \caption{ Invariant-mass distributions of (top left) $m_{p\Km}$, (top right) $m_{\antiproton\Kp}$, (bottom left) $m_{p\pim}$, and (bottom right) $m_{\antiproton\pip}$. Fit results are superimposed. Reproduced from~\cite{LHCb-PAPER-2018-025}. }
 \label{fig:Lb2ph}
\end{figure}

The LHCb analysis was performed using Run 1 data, with a total of about 9 000 $\Lb\to\proton \Km$ decays and 6 000 $\Lb\to\proton \pim$ decays.
The two-body invariant mass distributions and fit results used to obtain the raw yields are displayed in Figure~\ref{fig:Lb2ph}.
In addition, the difference in \CP asymmetries between $\Lb\to p \Km$ and $\Lb \to p\pi^-$ decays, which is insensitive to systematic uncertainties, is also reported to be
\begin{equation}
\Delta A_{\CP} \equiv A_{\CP}^{p\Km} - A_{\CP}^{p\pi^-}=0.014\pm0.022\pm0.010.
\end{equation}
Overall, no evidence of \CP violation has been observed in these decays. %AE: please check if there need an indent. % We don't need an indent here
\vspace{-6pt} 

%%%%5
\subsection{\CPV in Three-Body Decays}
\label{sec:three_body_decays}
\begin{figure}[H]
% \centering
 \includegraphics[width=0.48\textwidth]{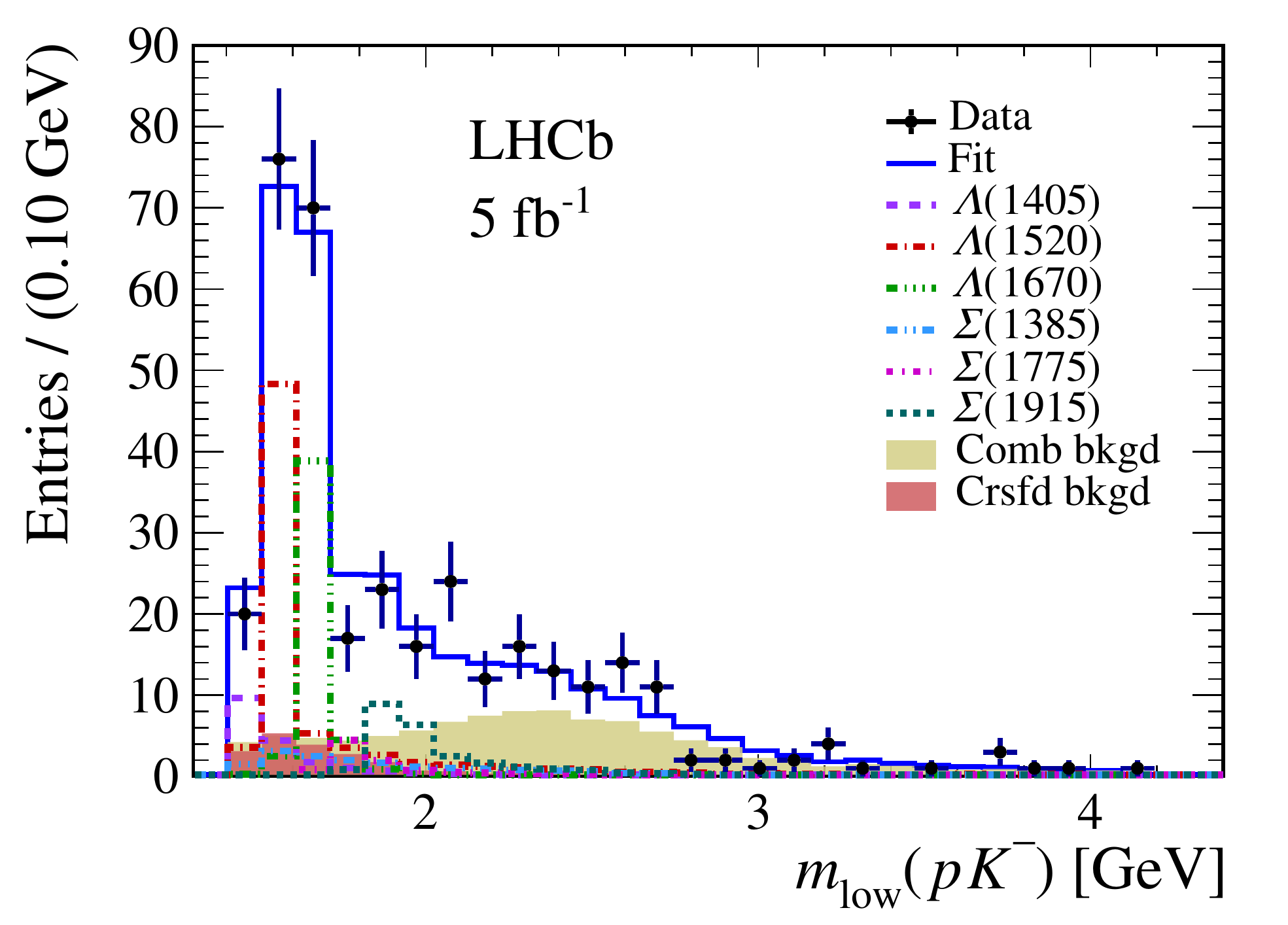}
 \includegraphics[width=0.48\textwidth]{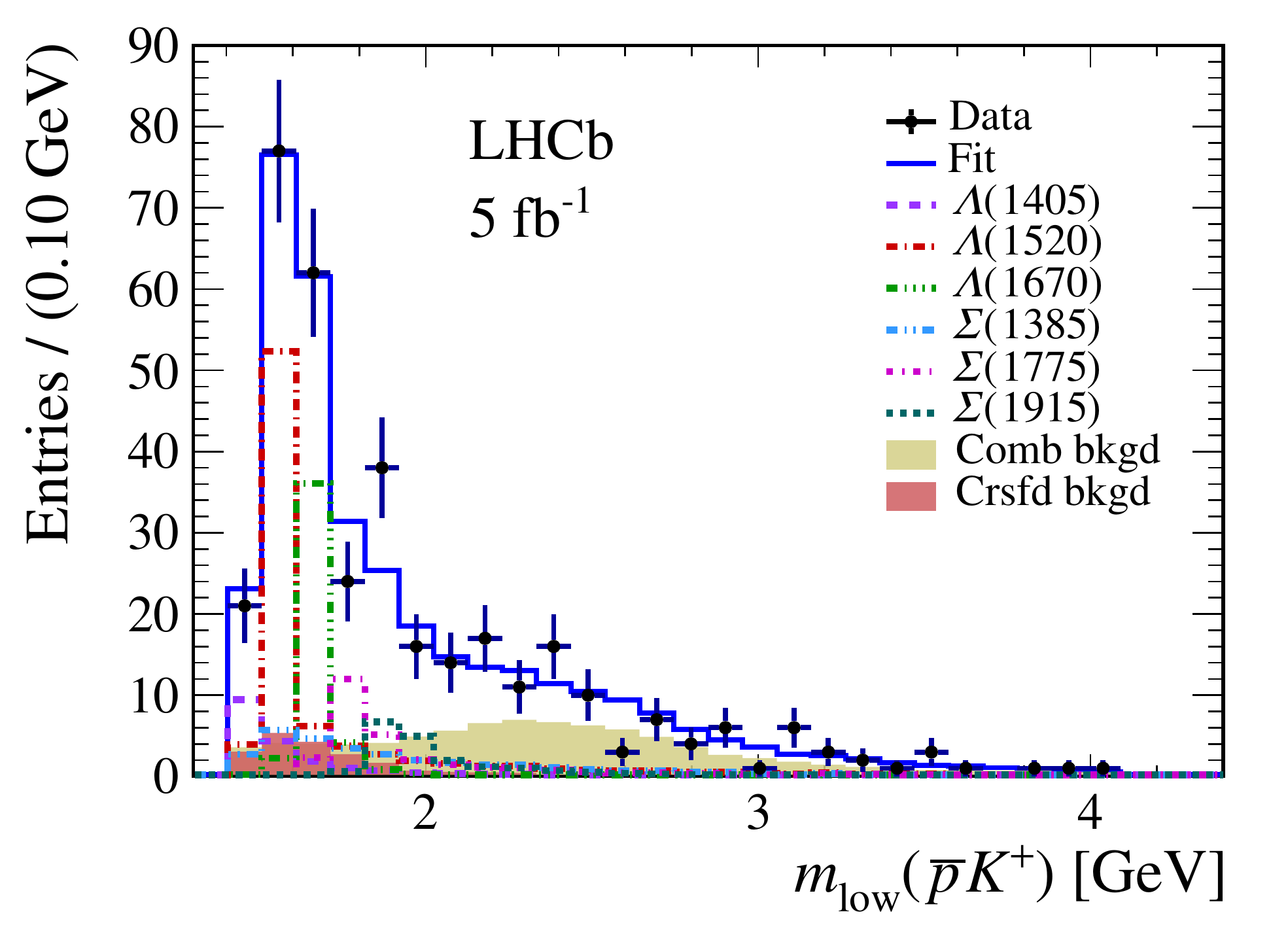}
 \caption{Distributions of the %MDPI: please change hyphen to minus sign in figure, such as change -1 to $-$1. % We prefer to keep -1 as they were in the the original article
 lower proton--kaon mass for (left) $\Xibm$ and (right) $\Xibbarp$ decays, with fit results superimposed. Reproduced from~\cite{LHCb-PAPER-2020-017}. }
 \label{fig:Xib2pKK}
\end{figure}

A search for \CP violation in the charmless decay $\Xibm \to p\Km\Km$ was performed using Run 1 and part of Run 2 data, resulting in a total of 500 signal events~\cite{LHCb-PAPER-2020-017}. The three-body final state was investigated using an amplitude analysis approach introduced in Section~\ref{sec:exp_methods}. A good description of the phase-space distribution was obtained with an amplitude model containing contributions from $\Sigma(1385)$, $\Lz(1405)^0$, $\Lz(1520)^0$, $\Lz(1670)^0$, $\Sigma(1775)^0$, and $\Sigma(1915)^0$ resonances, as demonstrated by Figure~\ref{fig:Xib2pKK}, where the $pK$-invariant-mass distributions are compared with the amplitude fit results. The \CP asymmetry for the contribution of resonance $R$ is defined as
\begin{equation}
 A_i^{\CP}=\frac{\int_{\Phi_3}\left(d \Gamma_R / d \Phi_3-d \overline{\Gamma}_R / d \Phi_3\right) d \Phi_3}{\int_{\Phi_3}\left(d \Gamma_R / d \Phi_3+d \overline{\Gamma}_R / d \Phi_3\right) d \Phi_3},
\end{equation}
where $\overset{(-)}{\Gamma}$ denotes the density of the final state phase-space for $\Xibm$ ($\Xibbarp$). It is equivalent to the definition in Equation~\eqref{eq:CPV_amp}, but sums over all the helicity-dependent sub-amplitudes of each resonance.
The results for these asymmetries are summarized in Table~\ref{tab:amplitude_para}, which were all found to be consistent with zero, given the limited precision.

\begin{table}[H] 
\caption{\small
 Results for the \CP asymmetries of $\Xibm\to\proton\Km\Km$ decays. Reproduced from~\cite{LHCb-PAPER-2020-017}.
}
\label{tab:amplitude_para}
%\centering

\setlength{\tabcolsep}{24.5mm}
\begin{tabular}{lr@{$\,\pm\,$}c@{$\,\pm\,$}l}
\toprule 
\textbf{Component} & \multicolumn{3}{c}{\textbf{\boldmath{$A^{\CP}$ ($10^{-2}$)}}}\\
\midrule 
$\Sigmares(1385)^0$ & $-27$ & 34 		& 73 \\
$\Lz(1405)^0$  & $ -1$ & 24 		& 32 \\
$\Lz(1520)^0$  & $ -5$ & \phantom{1}9 & \phantom{1}8 \\
$\Lz(1670)^0$  & $ 3$ & 14   & 10 \\
$\Sigmares(1775)^0$ & $-47$ & 26   	& 14 \\
$\Sigmares(1915)^0$ & $ 11$ & 26   	& 22 \\ [0.3ex]
\bottomrule
\end{tabular} 
\end{table}

\vspace{-12pt} 

%________________________

A study of $\Lb \to D p \Km$ decays with $D$ going to the $\Km \pip$ and $\Kp \pim$ final states was performed based on the full Run 1 and Run 2 data~\cite{LbToD0pk_nonresonant}. 
About 1400 signal decays were reconstructed for the $\Lb \to [\Km\pip]_D p \Km$ decay, which is dominated by the CKM favoured %please ensure that the original meaning is retained %Re: reverted to orignal text, which is correct. It is favoured instead of flavored
 $\Lb\to\Dz p\Km $ process, while the $\Lb \to [\Kp\pim]_D p \Km$ decay is contributed by $\Lb \to \Dz p \Km$ with $\Dz\to\Kp\pim$ and $\Lb \to \Dzb p \Km$ with $\Dzb\to\Kp\pim$ processes. They are of similar magnitude, and both involve (suppressed) off-diagonal CKM matrix elements, such that only around 240~signal decays are detected. The two interfering processes of the $\Lb \to [\Kp\pim]_D p \Km$ decay make it sensitive to the CKM $\gamma$ angle~\cite{Zhang:2021sit}.
The branching fraction of the favored decay $\Lb \to [\Km\pip]_D p \Km$ over that of the suppressed decay $\Lb \to [\Kp\pim]_D p \Km$
was measured to be $R=7.1 \pm 0.8 _{-0.3}^{+0.4}$. The global \CP asymmetry of the $\Lb \to [\Kp\pim]_D p \Km$ decay was measured according to Equation~\eqref{eq:a_cp_exp} to be $A_\CP=0.12 \pm 0.09 _{-0.03}^{+0.02} $. The relative branching fraction and \CP asymmetry were also measured in the $p\Km$-resonance-dominated phase-space region $M^2(p\Km)<5\gevgevcccc$ to be $R=8.6 \pm 1.5_{-0.3}^{+0.4}$ and $A_\CP=0.01 \pm 0.16_{-0.03}^{+0.02}$, respectively.
The relative branching fraction in the $p\Km$-resonance region is consistent with a rough estimate based on relevant CKM matrix elements. No hint of \CPV was observed given the limited statistics.

%________________________
\CPV studies have also been attempted for $b$-baryon decays with a $\KS$ or $\Lz$ hadron in the final state, which is not favored by the LHCb due to the lower detection efficiencies for long-lived neutral particles. Within the Run 1 data, about 200 signal decays were observed for the $\Lb\to\Lz\Kp\Km$ decay and 100 for the $\Lb\to\Lz\Kp\pim$ decay~\cite{LHCb-PAPER-2016-004}. \CP asymmetries were measured using the CKM favoured $\Lb \to \Lc(\Lz\pip)\pim$ decay as a control mode to cancel the production and most detection asymmetries.
The results integrated over the final state phase-space were 
\begin{equation}
\begin{aligned}
& A_{\CP}(\Lambda^0_b \to \Lz\Kp\pim)=-0.53\pm0.23\pm0.11, \\
& A_{\CP}(\Lambda^0_b \to \Lz\Kp\Km)=-0.28\pm0.10\pm0.07,
\end{aligned}
\end{equation}
both indicating a consistency with the \CP symmetry.
%_____________________

The direct $\CP$ asymmetry measurement for the $\Lb\to \KS p \pim$ decay is only available for the 2011 data, with a total of about 200 signal decays~\cite{LHCb-PAPER-2013-061}.
Similarly, the $\Lb \to \Lc(p\KS) \pim$ decay is used as a control mode to cancel the production and detection asymmetries, giving
\begin{equation}
A_{\CP}(\Lb\to \KS p \pim)=0.22\pm0.13\pm0.11,
\end{equation}
which is consistent with no \CPV.

%%%%%%%%%
\subsection{\CPV in Charmless Four-Body Decays}
\label{sec:four_body_decays}
\begin{figure}[H]
% \centering
 \includegraphics[width=0.45\textwidth]{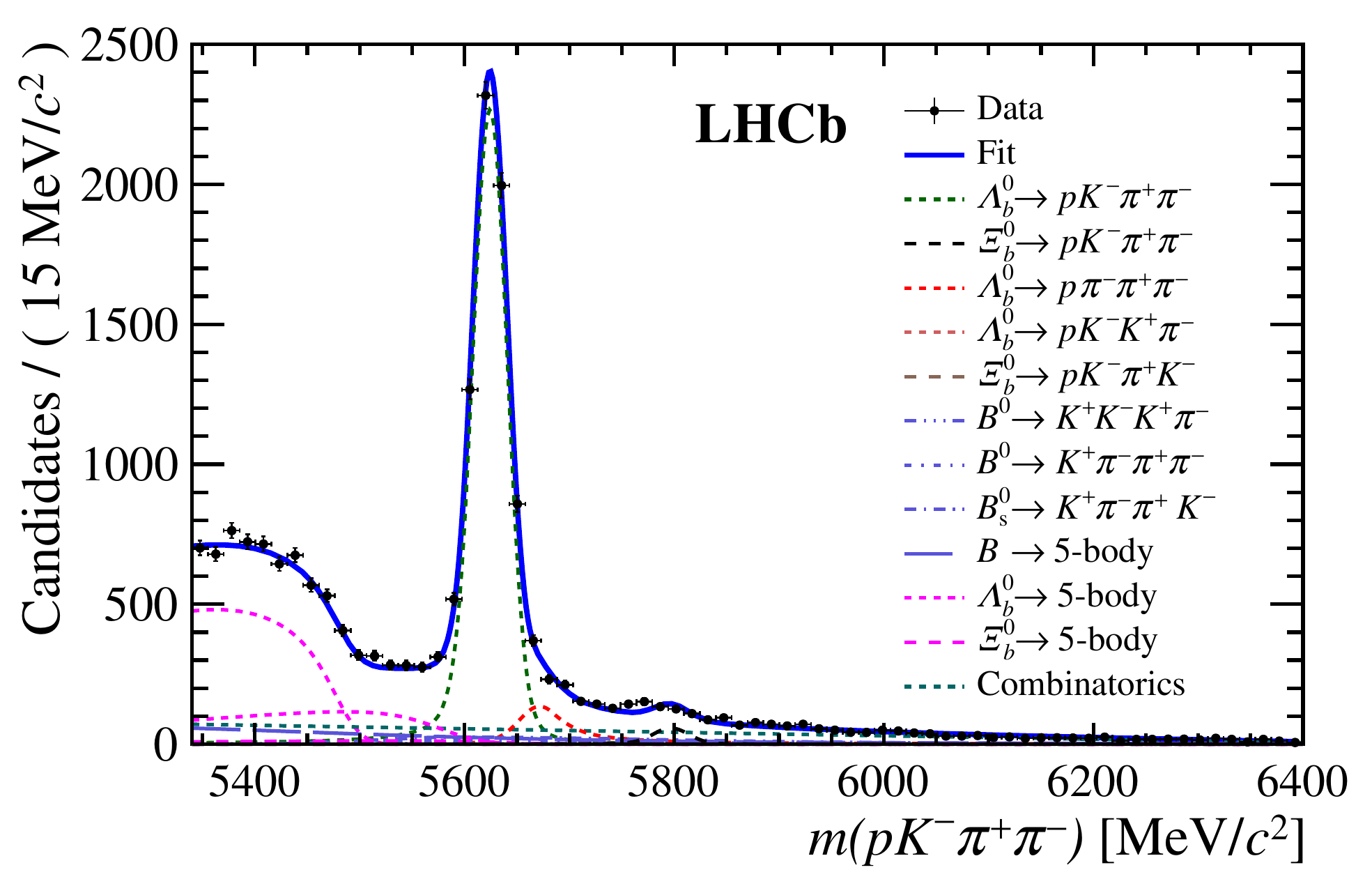}
 \includegraphics[width=0.45\textwidth]{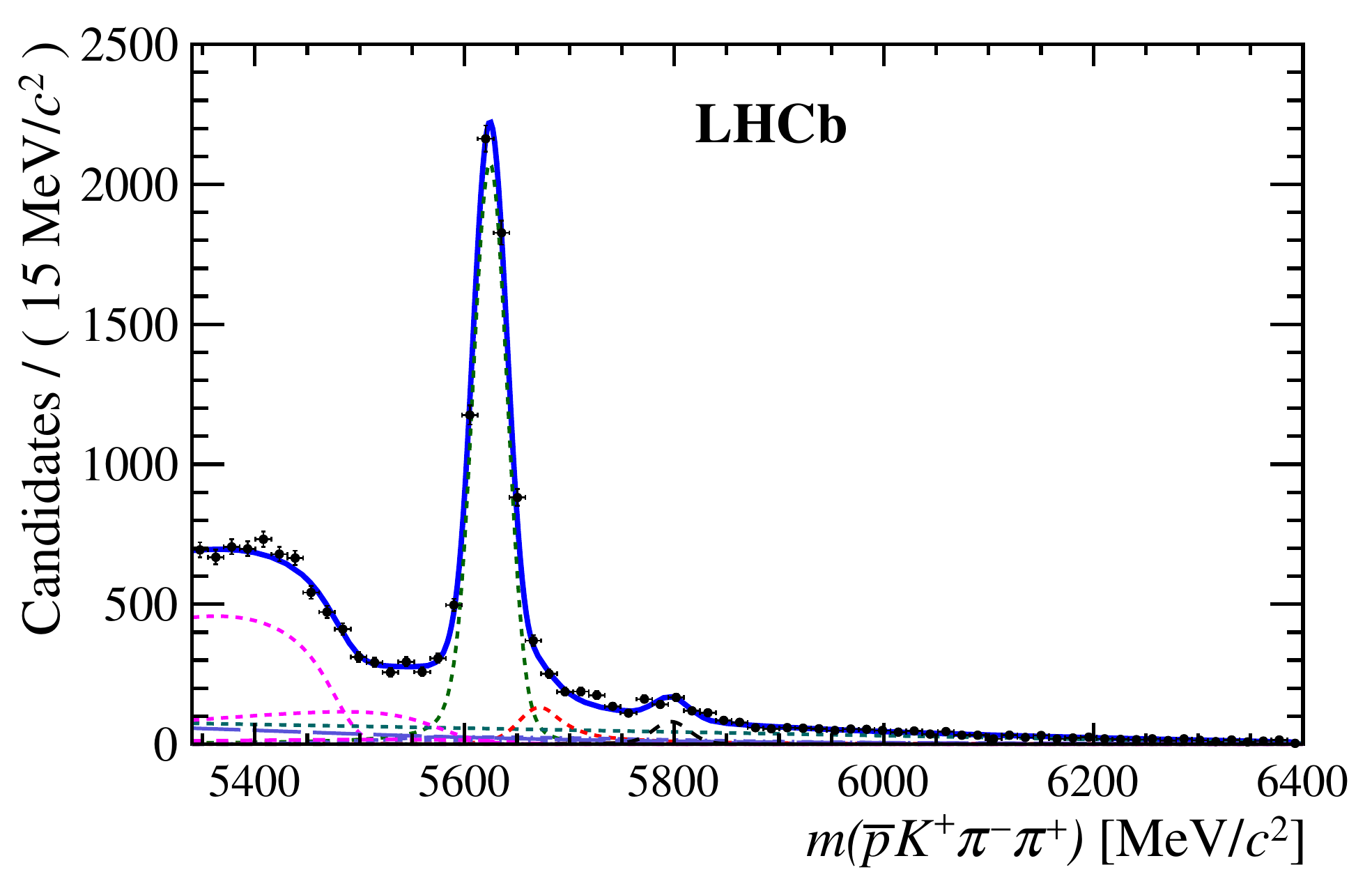}\\
 \includegraphics[width=0.45\textwidth]{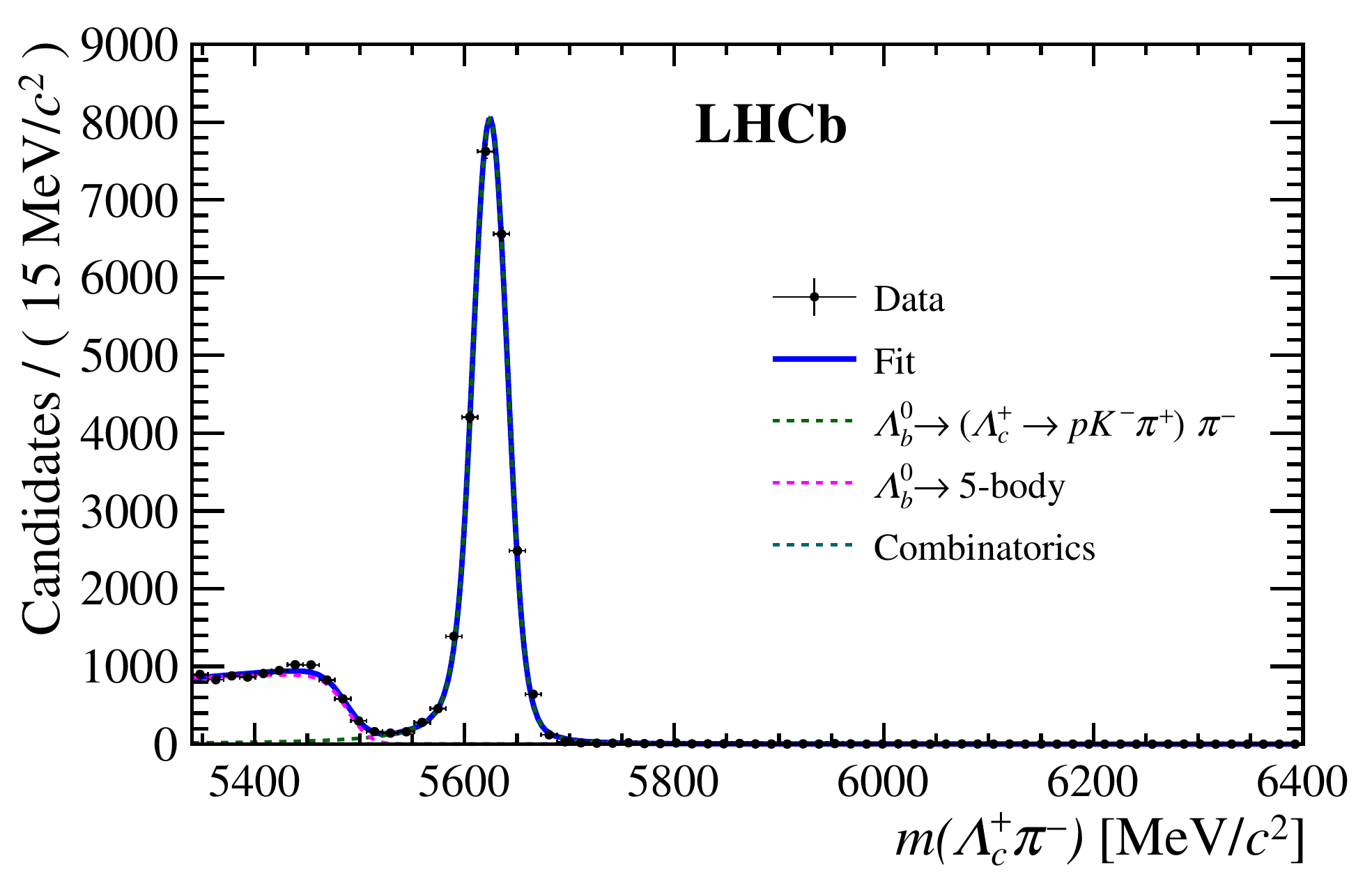}
 \includegraphics[width=0.45\textwidth]{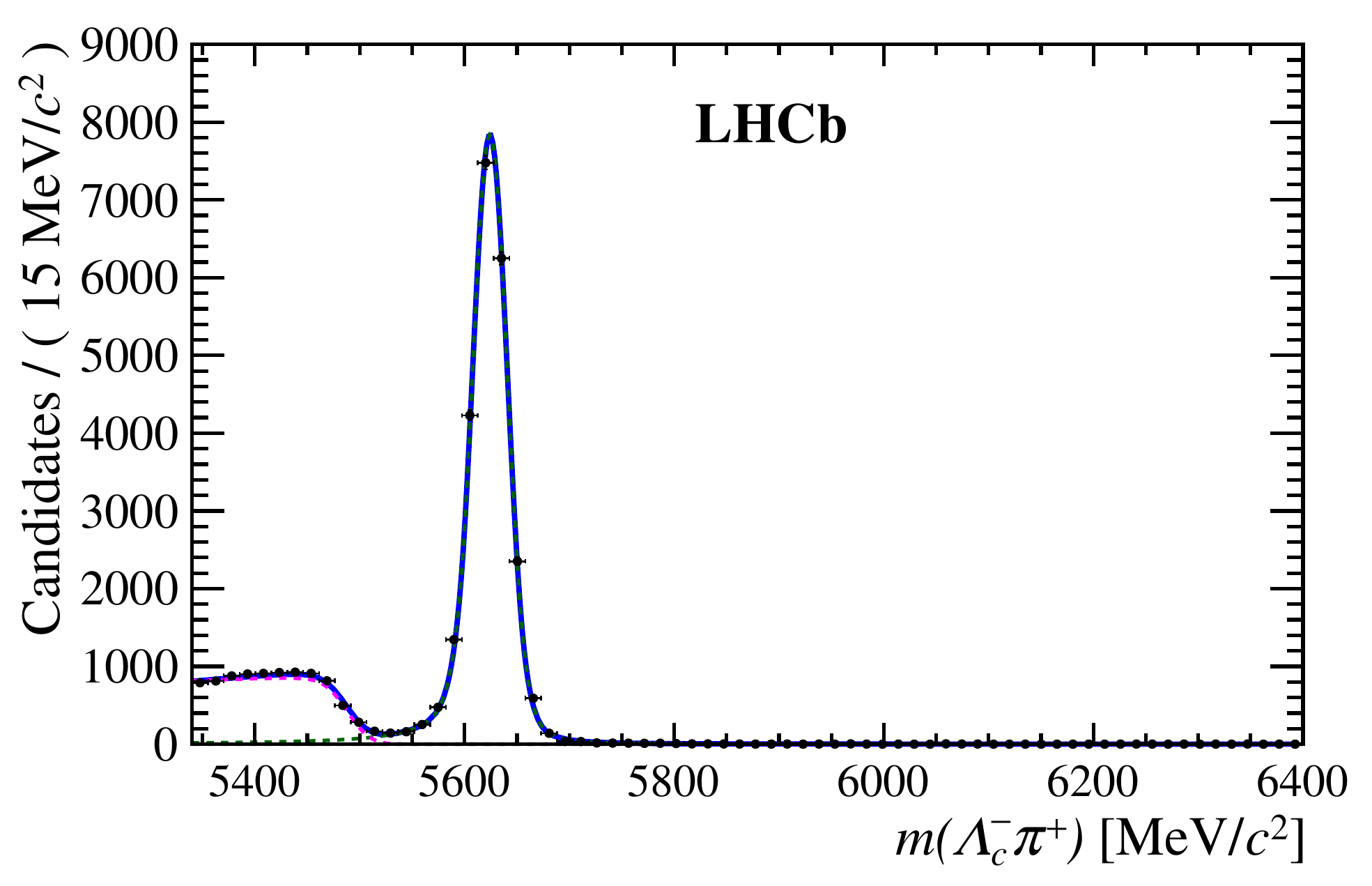}
 \caption{ Invariant-mass distributions of (\textbf{top left}) $m_{p \Km \pip \pim}$, (\textbf{top right}) $m_{\antiproton \Kp \pim \pip}$, (\textbf{bottom left}) $m_{\Lc \pim}$, and (\textbf{bottom right}) $m_{\Lambda_c^- \pip}$. Fit results are superimposed. {The legend from the top-left (\textbf{bottom-left}) figure is valid also for the top-right (\textbf{bottom-right}) figure.} Reproduced from~\cite{LHCb-PAPER-2018-044}. }
 \label{fig:Lb2pkhh}
\end{figure}

Charmless four-body decays are composed of abundant intermediate resonances that may receive contributions from tree and loop diagrams of varying magnitudes, causing the weak phase and the strong phase to vary in the final state phase-space. Significant localized $\CP$ asymmetries may be produced due to the interferences.
The LHCb collaboration studied \CP violation in charmless four-body decays of $\Lb$ and $\Xibz$ baryons, \mbox{$\Lb,\Xibz\to\proton h_1h_2h_3$}, using the Run 1 data~\cite{LHCb-PAPER-2018-044}. 
The production and detection asymmetries were canceled out by measuring \dacp with respect to the $\Lb\to\Lc\pim$ or $\Xibz\to\Xicp\pim$ control mode. The invariant mass distributions and fit results used to obtain raw yields of the $\Lb \rightarrow p \Km \pip \pim$ decay and its corresponding charmed decay $\Lb \rightarrow \Lc(p \Km \pip) \pim$ are displayed in Figure~\ref{fig:Lb2pkhh} as examples. A total of about 13 000 $\Lb \rightarrow p \Km \pip \pim$ charmless decays were observed.

In total, eighteen \CP asymmetries were measured, including the global \CP asymmetries for $\Lb$ and $\Xibz$ decays and localized asymmetries in the phase-space regions defined in Table~\ref{tab:massrequirement}, which put a focus on specific light resonances.

\begin{table}[H] 
\small
\caption{
Invariant-mass requirements %MDPI: we have revised lines in tables, please confirm. and please check if the alignment in table is correct, if no, please modify. %Re: everything is fine here
 used to select different phase-space regions for localized \CPV measurements. LBM refers to ``Low 2$\times$2-Body Mass''. Reproduced from~\cite{LHCb-PAPER-2018-044}.
}
\label{tab:massrequirement}
%\centering
\setlength{\tabcolsep}{5.2mm}
\begin{tabular}{lr}
\toprule
\textbf{Decay Mode} & \textbf{\boldmath{Invariant-Mass Requirements (in MeV/$c^2$)}} \\ \midrule
$\Lb \rightarrow p \pim \pip \pim$  & \multicolumn{1}{l}{} \\ \midrule
LBM & $m\left(p \pim \right)<2000\text { and } m\left(\pip \pim \right)<1640$ \\
$\Lb \rightarrow p a_1(1260)^{-}$  & $419<m\left(\pip \pim \pip \right)<1500$ \\
$\Lb \rightarrow N(1520)^0 \rho(770)^0$ & $1078<m\left(p \pim \right)<1800 \text { and } m\left(\pip \pim \right)<1100$ \\
$\Lb \rightarrow \Delta(1232)^{++} \pim \pim$ & $1078<m\left(p \pip\right)<1432$ \\ \midrule
$\Lb \rightarrow p \Km \pip \pim$ & \\ \midrule
LBM & $ m\left(p \Km\right)<2000 \text { and } m\left(\pip \pim \right)<1640 $ \\
$\Lb \rightarrow N(1520)^0 K^*(892)^0$ & $1078<m\left(p \pim \right)<1800 \text { and } 750<m\left(\pip \Km\right)<1100$ \\
$\Lb \rightarrow \Lambda(1520) \rho(770)^0$ & $1460<m\left(p \Km \right)<1580 \text { and } m\left(\pip \pim \right)<1100$ \\
$\Lb \rightarrow \Delta(1232)^{++} \Km \pim$ & $1078<m\left(p \pip \right)<1432$ \\
$\Lb \rightarrow p K_1(1410)^{-}$  & $1200<m\left(\Km \pip \pim\right)<1600$ \\ \midrule
$\Lb \rightarrow p \Km \Kp \Km$ & \multicolumn{1}{l}{} \\ \midrule
LBM & $m\left(p \Km\right)<2000 \text { and } m\left(\Kp \Km \right)<1675$ \\
$\Lb \rightarrow \Lambda(1520) \phi(1020)$ & $1460<m\left(p \Km \right)<1600 \text { and } 1005<m\left(\Kp \Km \right)<1040$ \\
$\Lb \rightarrow\left(p \Km \right)_{\text {high-mass }} \phi(1020)$ & $m\left(p \Km\right)>1600 \text { and } 1005<m\left(\Kp \Km\right)<1040$ \\ \bottomrule
\end{tabular}
\end{table}

The measurements of $\Delta \mathcal{A}^{\CP}$ in low two-body invariant-mass regions (``LBM'') (defined in Table~\ref{tab:massrequirement}) were: 
\begin{equation}
\begin{aligned}
\Delta \mathcal{A}^{\CP}\left(\Lb \rightarrow p \pim \pip \pim, \text{LBM}\right) & =(+3.7 \pm 4.1 \pm 0.5) \%, \\
\Delta \mathcal{A}^{\CP}\left(\Lb \rightarrow p \Km \pip \pim, \text{LBM}\right) & =(+3.5 \pm 1.5 \pm 0.5) \%, \\
\Delta \mathcal{A}^{\CP}\left(\Lb \rightarrow p \Km \Kp \Km, \text{LBM}\right) & =(+2.7 \pm 2.3 \pm 0.6) \%.
\end{aligned}
\end{equation}
The measurements for quasi two-body decays defined in Table~\ref{tab:massrequirement} were:
\begin{equation}
\begin{aligned}
\Delta \mathcal{A}^{\CP}\left(\Lb \rightarrow p a_1(1260)^{-}\right) & =(-1.5 \pm 4.2 \pm 0.6) \%, \\
\Delta \mathcal{A}^{\CP}\left(\Lb \rightarrow N(1520)^0 \rho(770)^0\right) & =(+2.0 \pm 4.9 \pm 0.4) \%, \\
\Delta \mathcal{A}^{\CP}\left(\Lb \rightarrow \Delta(1232)^{++} \pim \pim\right) & =(+0.1 \pm 3.2 \pm 0.6) \%, \\
\Delta \mathcal{A}^{\CP}\left(\Lb \rightarrow p K_1(1410)^{-}\right) & =(+4.7 \pm 3.5 \pm 0.8) \%, \\
\Delta \mathcal{A}^{\CP}\left(\Lb \rightarrow \Lambda(1520) \rho(770)^0\right) & =(+0.6 \pm 6.0 \pm 0.5) \%, \\
\Delta \mathcal{A}^{\CP}\left(\Lb \rightarrow N(1520)^0 K^*(892)^0\right) & =(+5.5 \pm 2.5 \pm 0.5) \%, \\
\Delta \mathcal{A}^{\CP}\left(\Lb \rightarrow \Delta(1232)^{++} \Km \pim\right) & =(+4.4 \pm 2.6 \pm 0.6) \%, \\
\Delta \mathcal{A}^{\CP}\left(\Lb \rightarrow \Lambda(1520) \phi(1020)\right) & =(+4.3 \pm 5.6 \pm 0.4) \%, \\
\Delta \mathcal{A}^{\CP}\left(\Lb \rightarrow\left(p \Km\right)_{\text {high-mass }} \phi(1020)\right) & =(-0.7 \pm 3.3 \pm 0.7) \% .
\end{aligned}
\end{equation}
The integrated (global) $\Delta \mathcal{A}^{\CP}$ were:
\begin{equation}
\begin{aligned}
\Delta \mathcal{A}^{\CP}\left(\Lb \rightarrow p \pim \pip \pim\right) & =(+1.1 \pm 2.5 \pm 0.6) \%, \\
\Delta \mathcal{A}^{\CP}\left(\Lb \rightarrow p \Km \pip \pim\right) & =(+3.2 \pm 1.1 \pm 0.6) \%, \\
\Delta \mathcal{A}^{\CP}\left(\Lb \rightarrow p \Km \Kp \pim\right) & =(-6.9 \pm 4.9 \pm 0.8) \%, \\
\Delta \mathcal{A}^{\CP}\left(\Lb \rightarrow p \Km \Kp \Km\right) & =(+0.2 \pm 1.8 \pm 0.6) \%, \\
\Delta \mathcal{A}^{\CP}\left(\Xibz \rightarrow p \Km \pip \pim\right) & =(-17 \pm 11 \pm 1) \%, \\
\Delta \mathcal{A}^{\CP}\left(\Xibz \rightarrow p \Km \pip \Km\right) & =(-6.8 \pm 8.0 \pm 0.8) \% .
\end{aligned}
\end{equation}

No significant \CP violating effect was observed in any of the measurements, ruling out \CPV as large as $\sim$10\%. The measurements are being updated to Run 2 data, which will improve the precision by more than a factor of two.

\begin{figure}[H]
% \centering
 \hspace{-6pt} \includegraphics[width=0.48\textwidth]{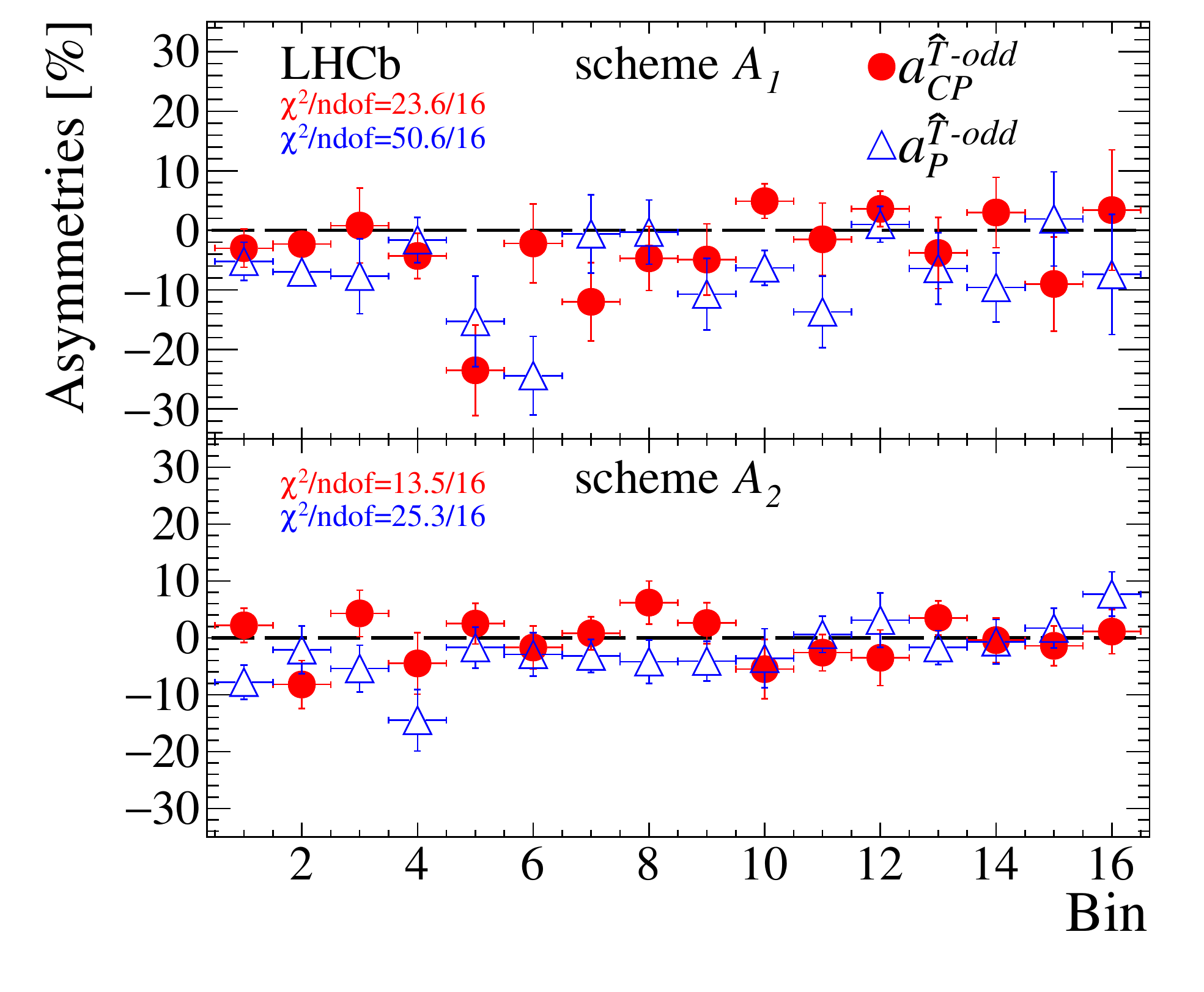}
 \includegraphics[width=0.48\textwidth]{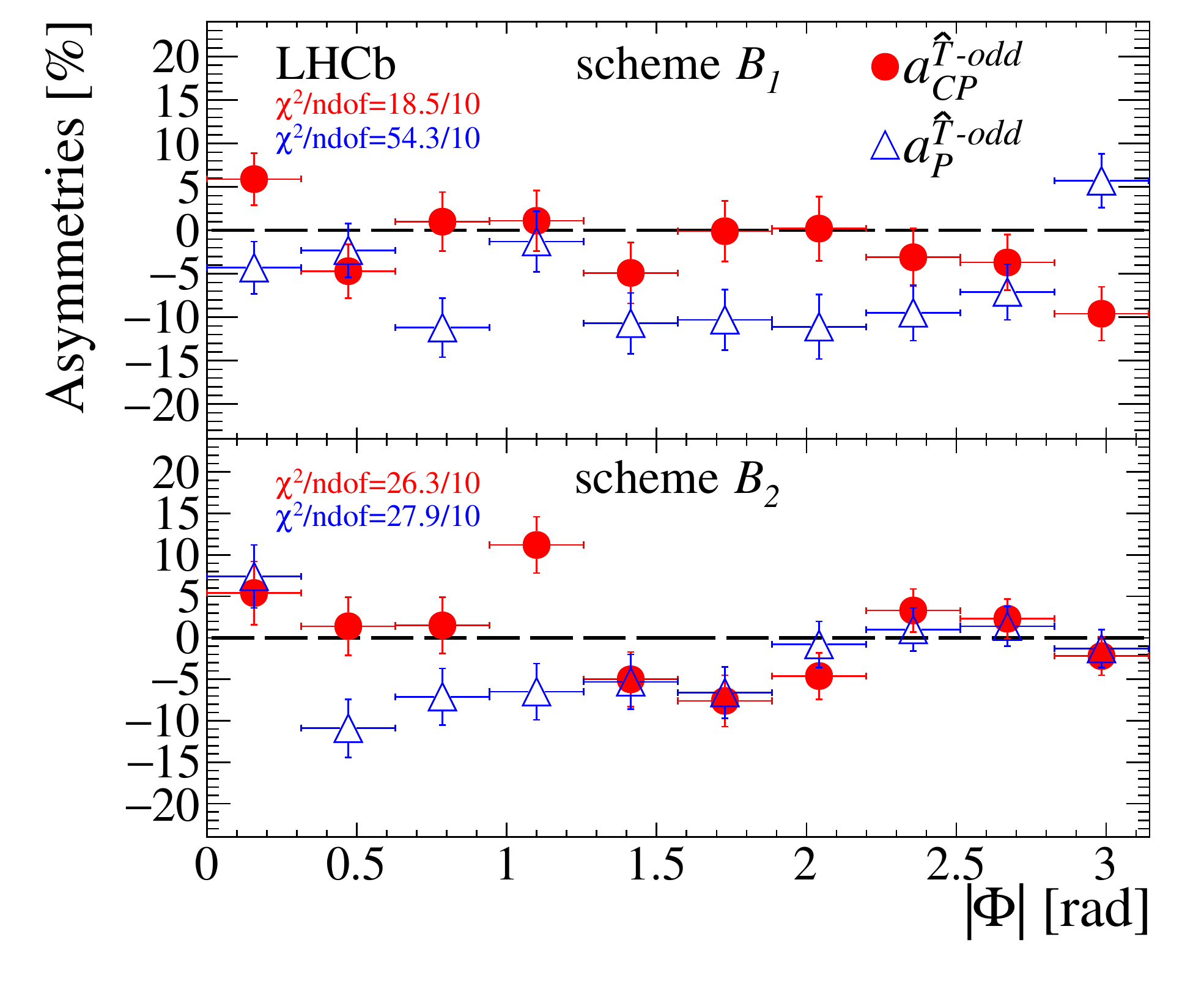}
 \vspace{-6pt} 
 \caption{ $\That$-odd $P$ and $\CP$ asymmetries of $\Lb\to\proton\pim\pip\pim$ decays measured for the binning scheme (left) $A_1$ and $A_2$ and (right) $B_1$ and $B_2$. Reproduced from~\cite{LHCb-PAPER-2019-028}. }
 \label{fig:Lb2p3pi}
\end{figure}

The TPA and the Energy Test methods discussed in Section~\ref{sec:exp_methods} were also exploited to study \CPV in $\Lb\to\proton\pim\pip\pim$ decays using data from Run 1 and 2015--2017~\cite{LHCb-PAPER-2019-028}. The dominant contributions proceed through the chain $\Lb\to N^{*+}\pim$, $N^{*+}\to\Delta(1234)^{++}\pim$ with $\Delta(1234)^{++}\to\proton\pip$ and the chain $\Lb\to a_1(1260)^-\pip$, $a_1(1260)^-\to \rho(770)^0\pim$ with $\rho(770)^0\to\pip\pim$.
With a total of about 27 000 signal decays, the global \CP asymmetry measured using the TPA method was $\aCPTodd=(-0.7\pm0.7\pm0.2)\%$. The $\aCPTodd$ parameter was also studied in final state phase-space regions defined according to polar and azimuthal angles of the proton in the $\Delta(1234)^{++}$ rest frame and the $\Delta(1234)^{++}$ in the $N^{*+}$ rest frame for the two $m(p\pip\pim_\text{slow})$-invariant-mass regions. The high mass region $m(p\pip\pim_\text{slow})>2.8\gevcc$ is dominated by the $a_1(1260)^-$ resonance, and the low one $m(p\pip\pim_\text{slow})<2.8\gevcc$ is dominated by the $N^{*+}$ resonance. These two groups of phase-space bins are referred to as $A_1$ and $A_2$, respectively. A second binning scheme is defined by dividing the absolute value of the angle ($|\Phi|$) between the planes of the $p\pim_\text{fast}$ system and the $\pip\pim_\text{slow}$ system, separately for the two $m(p\pip\pim_\text{slow})$ mass regions, referred to as $B_1$ and $B_2$, respectively. 
Here, $\pim_\text{fast}$ and $\pim_\text{slow}$ represent 
the faster and slower of two negative pions in the $\Lb$ rest frame, respectively. No evidence of \CPV was observed for any binning scheme, as shown in Figure~\ref{fig:Lb2p3pi}. For the study using the method of the Energy Test, where the distance is built from final state two-body and three-body invariant masses, no evidence of \CPV was found either. On the other hand, violation of parity $P$ was observed by more than 5.3 standard deviations with both methods, largely contributed by the high $m(p\pip\pim_\text{slow})$ mass, i.e., the $a_1(1260)^-$ region, as can be seen in Figure~\ref{fig:Lb2p3pi}.

\subsection{Rare Beauty Baryon Decays}
\label{sec:rare_decays}
\begin{figure}[H]
% \centering
 \includegraphics[width=0.48\textwidth]{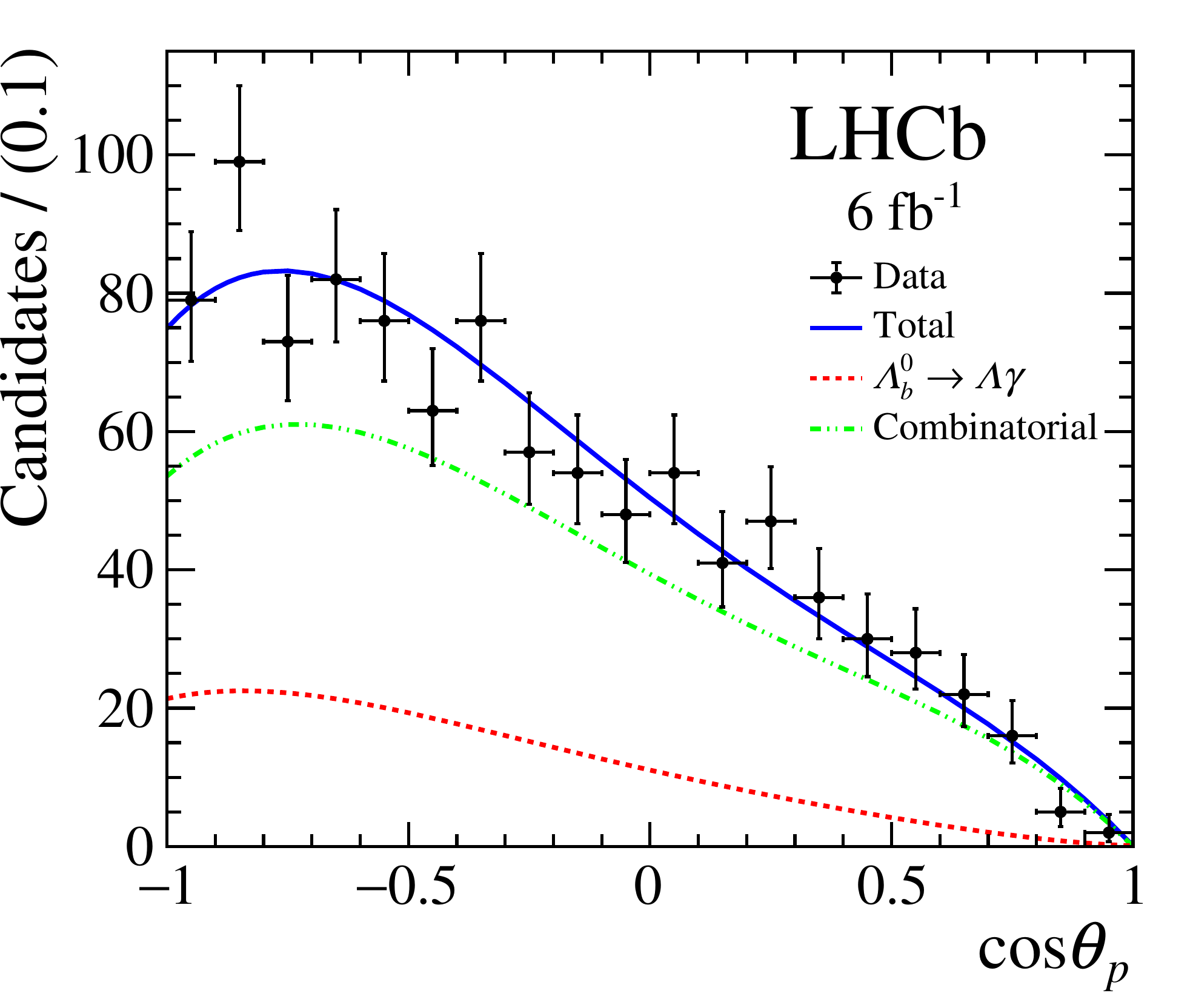}
 \includegraphics[width=0.48\textwidth]{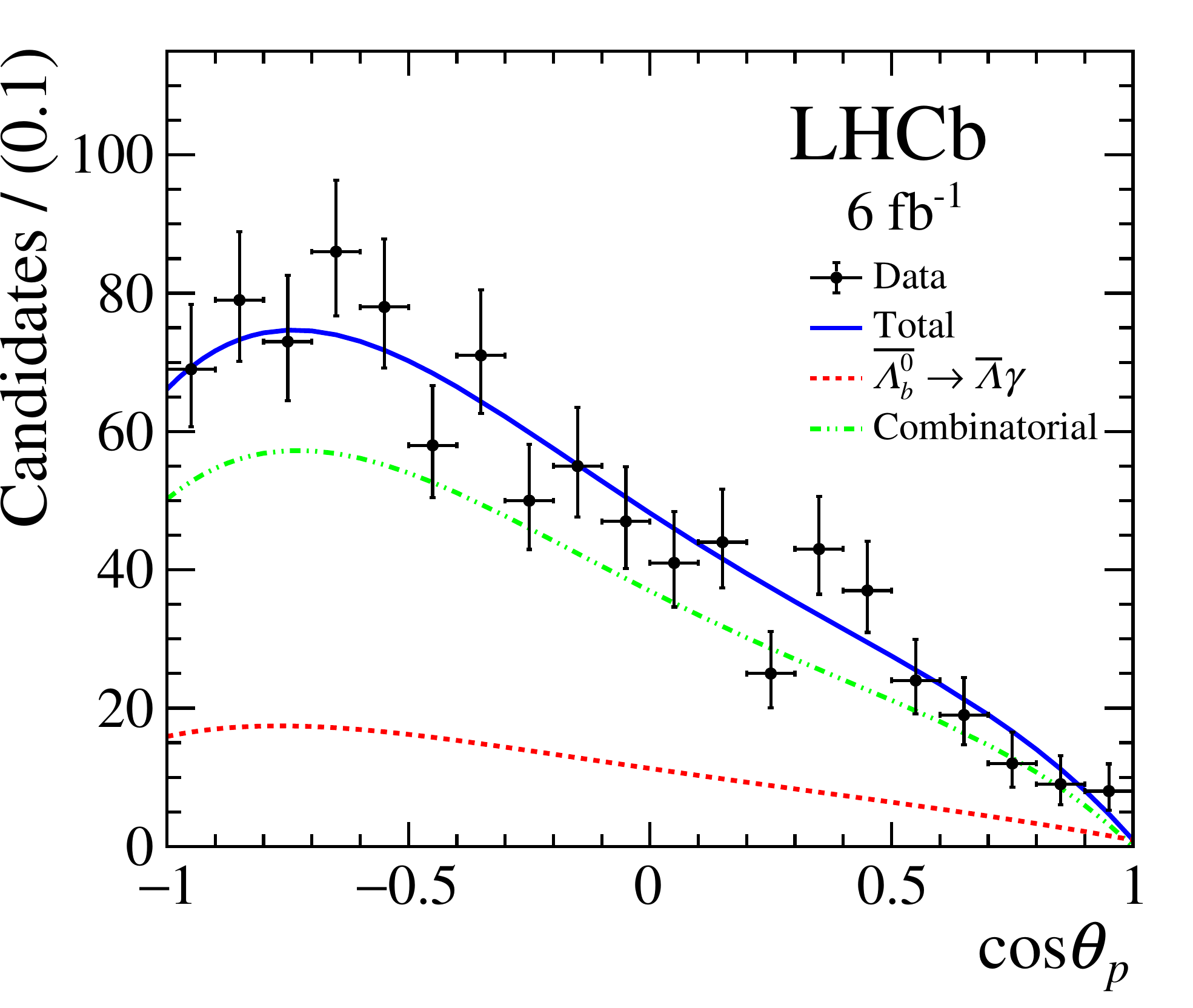}
 \caption{ The $\cos\theta_p$ %MDPI: please change hyphen to minus sign in figure, such as change -1 to $-$1. %Re: We prefer keep as they were in the orignal LHCb article
 distribution for (left) $\Lb\to\Lz\gamma$ and (right) $\Lbbar\to\Lbar\gamma$ candidates superimposed by the fit results. Reproduced from~\cite{LHCb-PAPER-2021-030}. }
 \label{fig:Lb2L0gamma}
\end{figure}

In the SM, Flavor-Changing Neutral Current (FCNC) decays are mediated only by loop diagrams, which are heavily suppressed, such that they are usually referred to as rare decays. For charmless decays of beauty hadrons, the FCNC diagrams interfere with the tree diagrams, which have different weak phases. The flavor structure of $b$-hadron FCNC decays is well known in the SM and is very sensitive to hypothetical new heavy particles in the loop.
The $\Lb \to \Lz \gamma$ decay, where $\Lz$ is reconstructed with $\Lz\to\proton\pim$, is contributed by the $b\to\squark\gamma$ FCNC diagram. The final state photon is almost fully left-handed, with the right-handed component suppressed by the ratio of strange-over-beauty quark masses $m_s^2/m_b^2$, such that parity $P$ is strongly violated. The $P$ asymmetry is quantified by the normalized difference, $\alpha_\gamma$, between the number of left-handed ($\gamma_L$) and right-handed ($\gamma_R$) photons, $\alpha_\gamma\equiv\frac{\gamma_L-\gamma_R}{\gamma_L+\gamma_R}$.
Dominated by a single diagram, \CPV is negligible for the $\Lb \to \Lambda \gamma$ decay.
The LHCb made the first measurement of $\alpha_\gamma$ for $\Lb \to \Lambda \gamma$ decays using the full Run 2 data with a total yield of about 450 signal decays~\cite{LHCb-PAPER-2021-030}. The parameter $\alpha_\gamma$ was extracted from an angular analysis to the proton polar angle ($\theta_p$) in the rest of $\Lz$, as illustrated in Figure~\ref{fig:Lb2L0gamma}. The results obtained separately for $\Lb$ and $\Lbbar$ decays were $\alpha_{\gamma}(\Lb) = 1.26\pm0.42\pm0.20$ and $\alpha_{\gamma}(\Lbbar) = -0.55\pm0.32\pm0.16$. These are consistent with a left(right)-handed $\Lb\to\Lz\gamma$ ($\Lbbar\to\Lbar\gamma$) decay and are compatible with no \CPV.

\begin{figure}[H]
% \centering
 \includegraphics[width=0.9\textwidth]{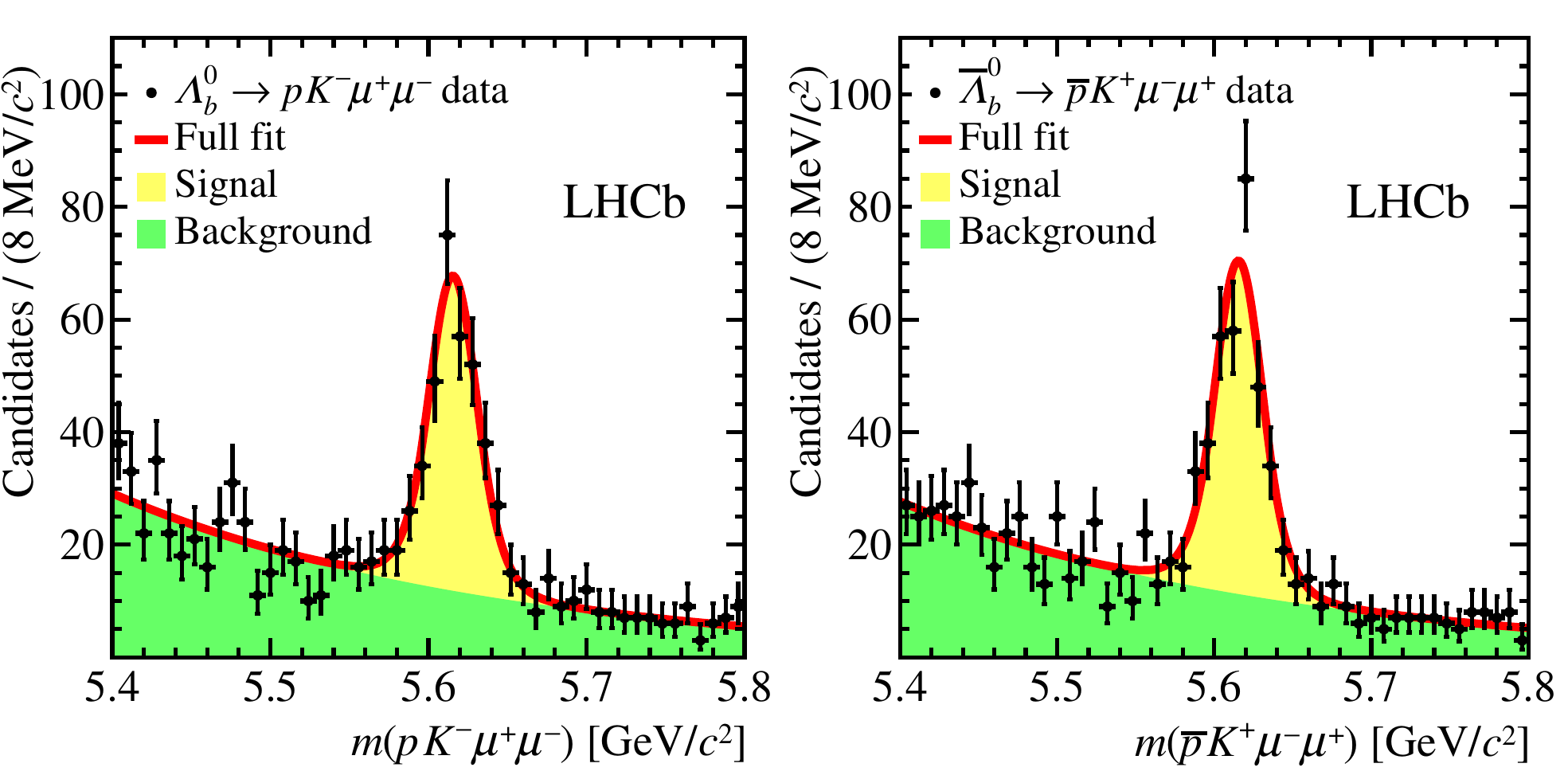}
 \caption{ Invariant mass distributions of (left) $\Lb\to p\Km \mumu$ and (right) $\Lbbar\to \antiproton \Kp \mumu$ decays, with fit results superimposed.
 Reproduced from~\cite{LHCB-PAPER-2016-059}. }
 \label{fig:Lb2pKmumu}
\end{figure}

The $\Lb\to \Lz^{(*)} \mu^+ \mu^-$ decay is an FCNC process with the underlying quark-level transition $b\to\squark\mumu$. At low $Q^2\equiv m^2(\mumu)$, it probes similar new physics effects as for $b\to\squark\gamma$ and provides additional information at high $Q^2$. In analogous $B\to K^{(*)}\mumu$ decays, $Q^2$-differential branching fractions and angular distributions have been found to have tensions with SM calculations~\cite{Chen:2021ftn}. These flavor anomalies have triggered much interest in the particle physics community. In the SM, all the dominating $b\to\squark\mumu$ diagrams have the same small weak phase, such that \CPV is limited, making it sensitive to \CP violation effects from physics beyond the SM.
The \CPV of the $\Lambda^0_b \to p \Km \mu^+ \mu^-$ decay is studied by the LHCb using the Run 1 data sample of around 600~signal decays~\cite{LHCB-PAPER-2016-059}, as shown by the invariant mass distributions in Figure~\ref{fig:Lb2pKmumu}. The direct \CP asymmetry is measured with respect to the control mode $\Lb\to \proton\Km\jpsi(\mumu)$, giving \mbox{$ A_{\CP}=(-3.5\pm5.0\pm0.2)\times 10^{-2}$}. The \CPV is also measured using the TPA method according to Equation~\eqref{eq:tpa_atp_atcp} as \mbox{$a^{\hat{T}-odd}_{\CP} = (1.2\pm5.0\pm1.7)\times 10^{-2}$}. Both are compatible with \CP conservation and agree with SM predictions~\cite{Paracha:2014hca,Alok:2011gv}.

\section{Results on Charm Baryon \boldmath{\CPV}}
\label{sec:charm_decays}
The charm sector is both complementary to measurements in $b$-hadrons and at the same time is unique for studies of \CP effects in decays of an up-type quark. Additionally, when compared to studies in the \textit{b}-sector, any theoretical predictions for the charm sector%please ensure that the original meaning is retained
 are more complicated due to the mass of charm baryons%please ensure that the original meaning is retained
, which is not negligible as in the case of light quarks, but also not heavy enough to use heavy quark expansion methods.

In the SM, \CP violation in the charm sector can only occur in Cabibbo-suppressed decays. The direct \CP asymmetry is expected to be at the $\mathcal{O}(10^{-3})$ level~\cite{Bigi:2012_Charm_baryon_CPV_expectations}. However, $SU(3)$ flavor breaking, rescattering, and new physics may enhance the \CP asymmetry to $\mathcal{O}(10^{-2})$. Although the SM does not produce \CP violation in Cabibbo-Favoured %please ensure that the original meaning is retained %Re: it should be  favoured rather than flavored. Changed it back 
 (CF) and Doubly Cabibbo-Suppressed (DCS) decays, it is still worthwhile to measure their \CP asymmetries. DCS decays provide a clean environment to find new physics due to the low rate of possible SM processes. CF decays are usually used as control modes to cancel instrumental~uncertainties. 

The LHCb collaboration studied the \CP asymmetries in decays of $\Lc \to p \Km \Kp$ and $\Lc \to p \pi^-\pi^+$ decays using the Run 1 data~\cite{LHCb-PAPER-2017-044}. The $\Lc$ baryon is produced in the semileptonic decay $\Lb \to \Lc \mu^- X$, where $X$ represents any additional particle(s). The presence of $\mun$ in the final state increases the trigger efficiency due to a highly sensitive muon detection at the \lhcb. $\Delta A_{\CP}\equiv A_{\CP}( p \Km \Kp)-A_{\CP}( p \pi^- \pi^+)$ is calculated to cancel the $\Lb$ production asymmetry, muon detection asymmetry, and $\Lc$ final state detection asymmetry, to be
\begin{equation}
\Delta A_{\CP} =(0.30\pm0.91\pm0.61)\%,
\end{equation}
which is consistent with \CP conservation. This result is dominated by the statistical uncertainty, and the \lhcb is currently working on an extension of this analysis using the Run 2 data and also including promptly produced \Lc decays.

\begin{figure}[H]
% \centering
 \includegraphics[width=0.8\textwidth]{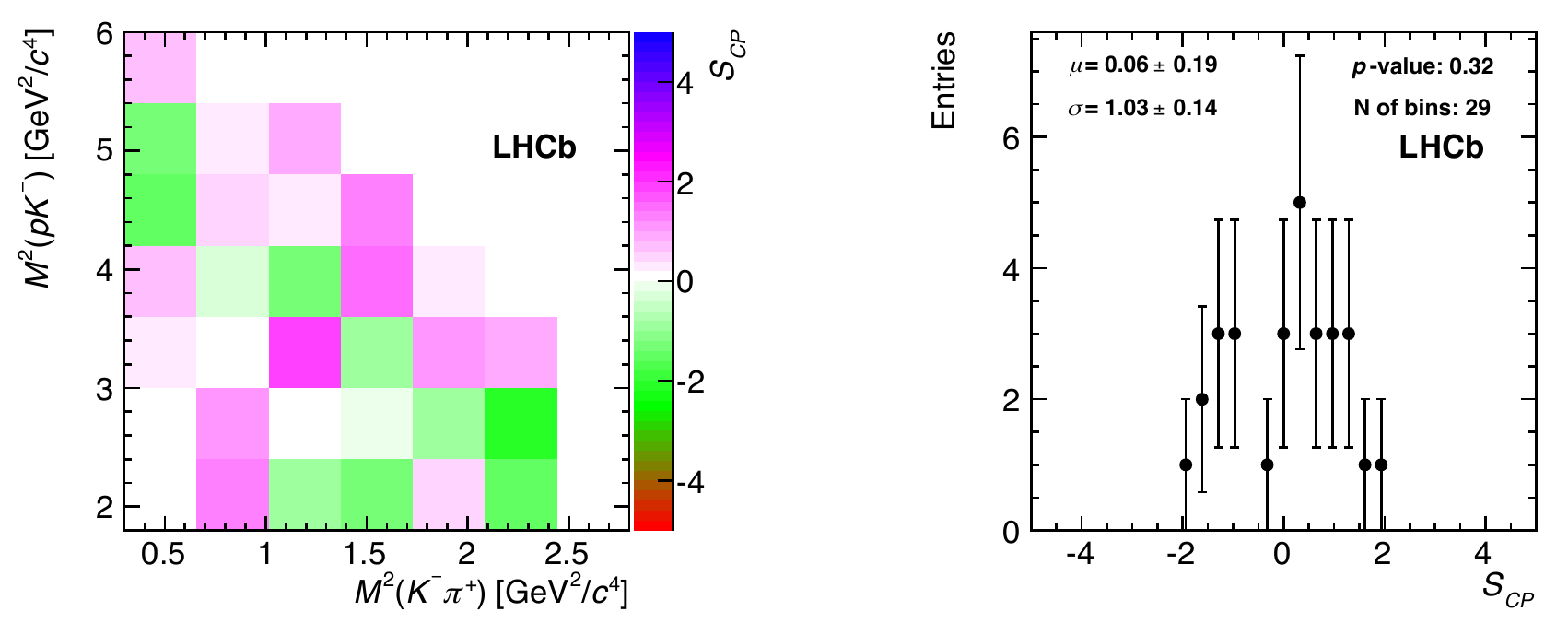}\\
 \includegraphics[width=0.8\textwidth]{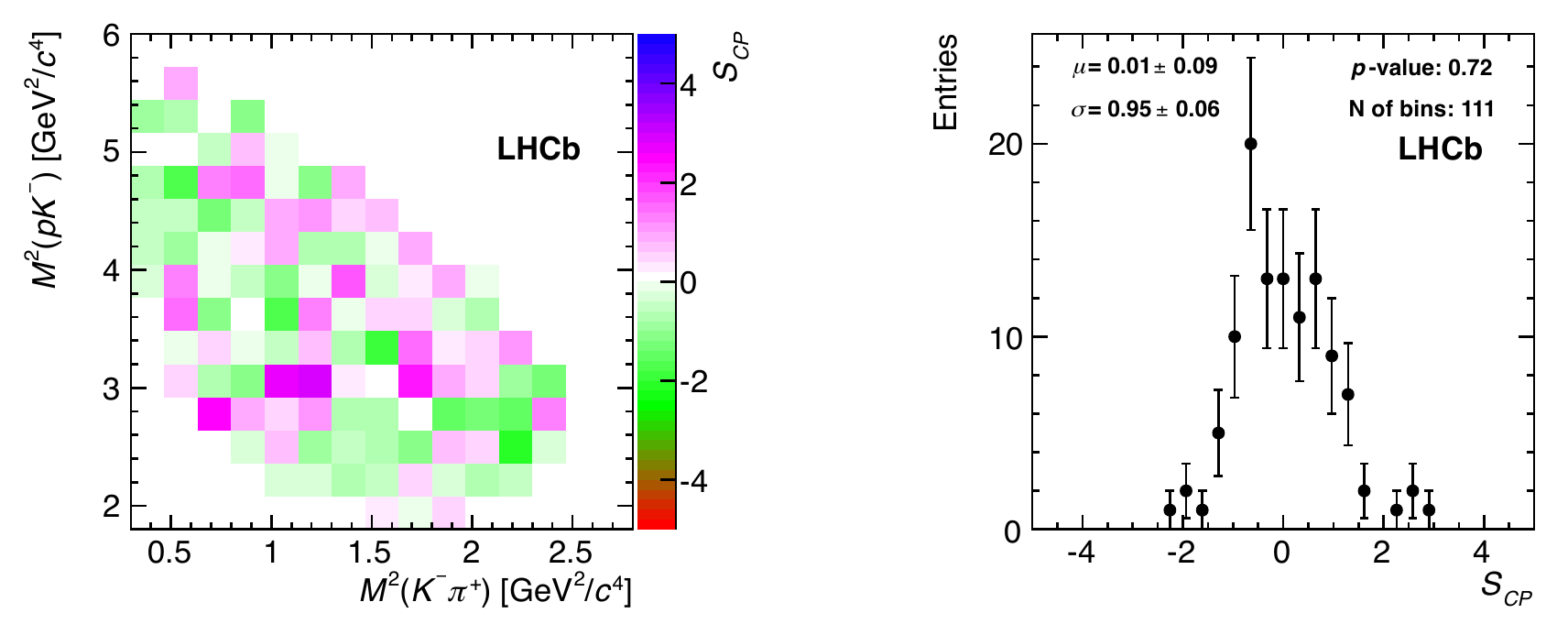}
 \caption{ Distributions %MDPI: please change hyphen to minus sign in figure, such as change -1 to $-$1. %Re: We prefer to keep as they were in the orignal LHCb article
 of $\mathcal{S}_{\CP}^i$ and corresponding one-dimensional distributions for $\Xicp \to \proton \Km \pip$ decays in the \lhcb Run 1 data: (top row) 29 uniform bins and (bottom row) 111 uniform bins of the three-body phase-space. Reproduced from~\cite{LHCB-PAPER-2019-026}. }
 \label{fig:Xic2pkpi}
\end{figure}

As discussed above, for multi-body decays, not only the global \CP asymmetry, but also localized asymmetries in the phase-space can be studied, which is the case for the measurement of \CPV for the Cabibbo-suppressed decay $\Xicp \to \proton \Km \pip$ using the Run 1 data~\cite{LHCB-PAPER-2019-026}.
In comparison to the above-described \Lc measurements, decays of promptly produced \Xicp are studied for this analysis.
The CF decay $\Lc \to p \Km \pip$ is used to subtract experimental asymmetries.
Due to the lack of knowledge about the $\Xi_c^+ \to p \Km \pi^+$ decay amplitude, two model-independent methods, the binned Miranda and the unbinned kNN, are employed to examine \CP asymmetries. The distribution of $S_{\CP}$ for the Miranda method shown in Figure~\ref{fig:Xic2pkpi} is consistent with a normal distribution, which corresponds to the \CP symmetry.
Similarly, the results obtained by using the kNN method are also consistent with the \CP~symmetry. 

\section{Summary and Prospects}
\label{sec:prospects}

The \lhcb collaboration is highly active in studies of \CP violation in baryon decays. This article reviewed recent results on \CPV and rare decays in beauty and charm baryons, with emphases on charmless two-body, three-body, and four-body beauty decays and Cabibbo-suppressed charm decays. In brief, no \CPV has been observed in baryon decays yet. It is noted that most of them have been obtained using the Run 1 or some combination with part of the Run 2 data. Studies on the most-promising channels using the full \lhcb dataset are ongoing.
At the same time, thanks to the largest recorded beauty and charm baryon datasets, the \lhcb can search for \CP violations in many additional decay modes.

\textls[-15]{One of the interesting opportunities is to measure the pseudo two-body decay \mbox{$\Lb \to \Lz D$}}, where $D\equiv\Dz+\Dzb$. A large \CP asymmetry is expected within the CKM model, arising from the interference between the $\bquark \to \squark + \Dz$ and $\bquark \to \squark + \Dzb$ amplitude, governed by \Vcb\Vuss and \Vub\Vcss CKM matrix elements, respectively. The decay is complementary to $B$-meson modes for determinations of the CKM angle $\gamma$. With the Run 1 and Run 2 data, the sensitivity to the angle gamma has been estimated to be $12$--$36^{\circ}$~\cite{Zhang:2021sit}.
This study is expected to benefit greatly from Run 3 of the LHCb, where the expected sensitivity for the angle $\gamma$ could reach $4$--$11^{\circ}$.

Another large area of potentially interesting decays are charmless decays of \mbox{\Xibm \to \Lz \hadron \hadron \hadron}, \mbox{\Xibm \to \proton \KS \hadron \hadron}, where \hadron represents a kaon or a pion, and similar decays resulting in five and more charged final state particles. These decays offer a very rich decay dynamics; however, their low production rate and low reconstruction efficiency arising from many final state particles make these studies challenging. 

With a new flexible software trigger and upgraded hardware, the \lhcb is in a good position to study all already discussed decays and many more with a significantly higher precision, allowing us to shed more light on the baryonic \CP violation puzzle. We look forward to more results from the LHCb in the coming years.

%%%%%%%%%%%%%%%%%%%%%%%%%%%%%%%%%%%%%%%%%%

%%%%%%%%%

%%%%%%%%%%%%%%%%%%%%%%%%%%%%%%%%%%%%%%%%%%
\vspace{6pt} 

%%%%%%%%%%%%%%%%%%%%%%%%%%%%%%%%%%%%%%%%%%
%% optional
%\supplementary{The following supporting information can be downloaded at: \linksupplementary{s1}, Figure S1: title; Table S1: title; Video S1: title.}


%%%%%%%%%%%%%%%%%%%%%%%%%%%%%%%%%%%%%%%%%%

\authorcontributions{Conceptualization, X.D., M.S., Y.S., X.Y., and Y.Z.; methodology, X.D., M.S., Y.S., X.Y., and Y.Z.; writing---original draft preparation, X.D., M.S., Y.S., X.Y., and Y.Z.; writing---review and editing, X.D., M.S., Y.S., X.Y., and Y.Z.; supervision, X.D., M.S., Y.S., X.Y., and Y.Z.; project administration, X.D., M.S., Y.S., X.Y., and Y.Z.; funding acquisition, X.D., M.S., Y.S., X.Y., and Y.Z. All authors have read and agreed to the published version of the manuscript.}

\funding{This research was funded by the National Natural Science Foundation of China (NSFC) under Contract Nos. 12061141007 and 12175005 and by the National Key Research and Development Program of China under Contract No. 2022YFA1601904. M.S. and Y.Z. were funded by Peking University by the Boya program.}

\dataavailability{Not applicable.} 

\acknowledgments{This work was supported by the State Key Laboratory of Nuclear Physics and Technology.}

\conflictsofinterest{The authors declare no conflict of interest.} 

%%%%%%%%%%%%%%%%%%%%%%%%%%%%%%%%%%%%%%%%%%
%% Optional
\sampleavailability{Not applicable.}

%% Only for journal Encyclopedia
%\entrylink{The Link to this entry published on the encyclopedia platform.}

\abbreviations{Abbreviations}{
The following abbreviations are used in this manuscript:\\

\noindent 
\begin{tabular}{@{}ll}
$P$ & Parity\\
$C$ & Charge conjugation\\
$CP$ & Charge conjugation and Parity\\
$CPV$ & $CP$ Violation\\
SM & Standard Model\\
CKM & Cabibbo--Kobayashi--Maskawa\\
FCNC & Flavor-Changing Neutral Current\\
DCS & Doubly Cabibbo Suppressed\\
CF & Cabibbo Favoured%please ensure that the original meaning is retained %Re: should be Favoured rather than flavored
\\
BAU & Baryon Asymmetry of the Universe\\
kNN & $k$-Nearest Neighbors\\
TPA & Triple Product Asymmetry\\
LHC &Large Hadron Collider\\
$LHCb$ & Large Hadron Collider Beauty Experiment\\
VELO & VErtex LOcator \\
TT & Tracker Turicensis \\
ECAL & Electromagnetic Calorimeter \\
HCAL & Hadronic Calorimeter
\end{tabular}
}

%%%%%%%%%%%%%%%%%%%%%%%%%%%%%%%%%%%%%%%%%%
\begin{adjustwidth}{-\extralength}{0cm}

\reftitle{References}

%\PublishersNote{}
\PublishersNote{}
\end{adjustwidth}


\begin{thebibliography}{999}

\bibitem[Noether(1918)]{Noether:1918zz}
Noether, E.
\newblock {Invariant Variation Problems}.
\newblock {\em Gott. Nachr.} {\bf 1918}, {\em 1918},~235--257.
 \href{http://xxx.lanl.gov/abs/physics/0503066}
\newblock
 {\changeurlcolor{black}\href{https://doi.org/10.1080/00411457108231446}{\detokenize{https://doi.org/10.1080/00411457108231446}}}.

\bibitem[Lee and Yang(1956)]{Lee:1956qn}
Lee, T.D.; Yang, C.N.
\newblock {Question of Parity Conservation in Weak Interactions}.
\newblock {\em Phys. Rev.} {\bf 1956}, {\em 104},~254--258.
\linebreak \newblock
 {\changeurlcolor{black}\href{https://doi.org/10.1103/PhysRev.104.254}{\detokenize{https://doi.org/10.1103/PhysRev.104.254}}}.

\bibitem[Wu \em{et~al.}(1957)Wu, Ambler, Hayward, Hoppes, and
 Hudson]{Wu:1957my}
Wu, C.S.; Ambler, E.; Hayward, R.W.; Hoppes, D.D.; Hudson, R.P.
\newblock {Experimental Test of Parity Conservation in Beta Decay}.
\newblock {\em Phys. Rev.} {\bf 1957}, {\em 105},~1413--1414.
\newblock
 {\changeurlcolor{black}\href{https://doi.org/10.1103/PhysRev.105.1413}{\detokenize{https://doi.org/10.1103/PhysRev.105.1413}}}.

\bibitem[Christenson \em{et~al.}(1964)Christenson, Cronin, Fitch, and
 Turlay]{Cronin:1964fg}
Christenson, J.H.; Cronin, J.W.; Fitch, V.L.; Turlay, R.
\newblock {Evidence for the $2\pi$ Decay of the $K_2^0$ Meson}.
\newblock {\em Phys. Rev. Lett.} {\bf 1964}, {\em 13},~138--140.
\newblock
 {\changeurlcolor{black}\href{https://doi.org/10.1103/PhysRevLett.13.138}{\detokenize{https://doi.org/10.1103/PhysRevLett.13.138}}}.

\bibitem[Kobayashi and Maskawa(1973)]{Kobayashi:1973fv}
Kobayashi, M.; Maskawa, T.
\newblock {CP Violation in the Renormalizable Theory of Weak Interaction}.
\newblock {\em Prog. Theor. Phys.} {\bf 1973}, {\em 49},~652--657.
\newblock
 {\changeurlcolor{black}\href{https://doi.org/10.1143/PTP.49.652}{\detokenize{https://doi.org/10.1143/PTP.49.652}}}.

\bibitem[Sakharov(1967)]{Sakharov:1967dj}
Sakharov, A.D.
\newblock {Violation of CP Invariance, C asymmetry, and baryon asymmetry of the
 universe}.
\newblock {\em Pisma Zh. Eksp. Teor. Fiz.} {\bf 1967}, {\em 5},~32--35.
\newblock
 {\changeurlcolor{black}\href{https://doi.org/10.1070/PU1991v034n05ABEH002497}{\detokenize{https://doi.org/10.1070/PU1991v034n05ABEH002497}}}.

\bibitem[Aubert \em{et~al.}(2002)Aubert et~al.]{BaBar:2001yhh}
Aubert, B.; Bazan, A.; Boucham, A.; Boutigny, D.; Bonis, I.D.; Favier, J.; Gaillard, J.-M.; Jeremie, A.; Karyotakis, Y.; Flour, T.L.; et~al.
\newblock {The BaBar detector}.
\newblock {\em Nucl. Instrum. Meth. A} {\bf 2002}, {\em 479},~1--116.
 \href{http://xxx.lanl.gov/abs/hep-ex/0105044}
\newblock
 {\changeurlcolor{black}\href{https://doi.org/10.1016/S0168-9002(01)02012-5}{\detokenize{https://doi.org/10.1016/S0168-9002(01)02012-5}}}.

\bibitem[Abashian \em{et~al.}(2002)Abashian et~al.]{Belle:2000cnh}
Abashian, A.; Gotow, K.; Morgan, N.; Piilonen, L.; Schrenk, S.; Abe, K.; Adachi, I.; Alexander, J.P.; Aoki, K.; Behari, S.; et~al.
\newblock {The Belle Detector}.
\newblock {\em Nucl. Instrum. Meth. A} {\bf 2002}, {\em 479},~117--232.
\newblock
 {\changeurlcolor{black}\href{https://doi.org/10.1016/S0168-9002(01)02013-7}{\detokenize{https://doi.org/10.1016/S0168-9002(01)02013-7}}}.

\bibitem[Alves \em{et~al.}(2008)Alves et~al.]{LHCb_detector}
Alves, A.A.; Filho, L.M.A.; Barbosa, A.F.; Bediaga, I.; Cernicchiaro, G.; Guerrer, G.; Lima, H.P., Jr.; Machado, A.A.; Magnin, J.; Marujo, F.; et al
\newblock {The LHCb Detector at the LHC %MDPI: we have removed extra information, please confirm.%Re: the update is ok
}.
\newblock {\em JINST} {\bf 2008}, {\em 3},~S08005.
 {\changeurlcolor{black}\href{https://doi.org/10.1088/1748-0221/3/08/S08005}{\detokenize{https://doi.org/10.1088/1748-0221/3/08/S08005}}}.

\bibitem[Amhis \em{et~al.}(2022)Amhis et~al.]{HFLAV:2022pwe}
Amhis, Y., Banerjee, S., Ben-Haim, E., Bertholet, E., Bernlochner, F. U., Bona, M.; Bozzi, C.; Brodzicka, J.; Chrzaszcz, M.; Dingfelder, J.; et al.
\newblock {Averages of $b$-hadron, $c$-hadron, and $\tau$-lepton properties as
 of 2021} \emph{ Eur. Phys. J. C} {\bf 2022}, \emph{81}, 226.


\bibitem[Aaij \em{et~al.}(2015)Aaij et~al.]{LHCb_perf_Run1}
The LHCb Collaboration.
\newblock {LHCb Detector Performance}.
\newblock {\em Int. J. Mod. Phys. A} {\bf 2015}, {\em 30},~1530022.
 \href{http://xxx.lanl.gov/abs/1412.6352}
\newblock
\linebreak {\changeurlcolor{black}\href{https://doi.org/10.1142/S0217751X15300227}{\detokenize{https://doi.org/10.1142/S0217751X15300227}}}.

\bibitem[Aaij \em{et~al.}(2019)Aaij et~al.]{LHCb_perf_and_trigger_Run2}
Aaij, R.; Akar, S.; Albrecht, J.; Alexander, M.; Albero, A.A.; Amerio, S.; Anderlini, L.; D'Argent, P.; Baranov, A.; Barter, W.; et al.
\newblock {Design and performance of the LHCb trigger and full real-time
 reconstruction in Run 2 of the LHC}.
\newblock {\em JINST} {\bf 2019}, {\em 14},~P04013.
 \href{http://xxx.lanl.gov/abs/1812.10790}
\newblock
 {\changeurlcolor{black}\href{https://doi.org/10.1088/1748-0221/14/04/P04013}{\detokenize{https://doi.org/10.1088/1748-0221/14/04/P04013}}}.

\bibitem[Bediaga \em{et~al.}(2012)Bediaga et~al.]{LHCb_upgrade_TDR}
Bediaga, I.; Chanal, H.; Hopchev, P.; Cadeddu, S.; Stoica, S.; Calvo Gomez, M.; dos Reis, A.C.; Amato, S.; Carvalho Akiba, K.; De Paula, L.; et al.
\newblock {\emph{Framework TDR for the LHCb Upgrade: Technical Design Report}};
\newblock Technical report; The LHC experiments Committee: Geneva, Switzerland, %Newly added information, please confirm. %Re: it is ok
 2012.

\bibitem[LHC(2014)]{LHCb_trigger_TDR}
Miriam, C.G.; Xavier, V.C.; LHCb Collaboration. {\emph{LHCb Trigger and Online Upgrade Technical Design Report}};
\newblock Technical report; CERN: Geneva, Switzerland, 2014. %Re: it is ok

\bibitem[Abe \em{et~al.}(2010)Abe et~al.]{Belle-II:2010dht}
Abe, T.; Adachi, I.; Adamczyk, K.; Ahn, S.; Aihara, H.; Akai, K.; Aloi, M.; Andricek, L.; Aoki, K.; Arai, Y.; et al.
\newblock {Belle II Technical Design Report}. \emph{arXiv} {\bf 2010}, arXiv:1011.0352.


\bibitem[Aaij \em{et~al.}(2021)Aaij et~al.]{LHCb:2021xyh}
The LHCb Collaboration; Aaij, R.; Abdelmotteleb, A.S.W.; Beteta, C.A.; Ackernley, T.; Adeva, B.; Adinolfi, M.; Afsharnia, H.; Aidala, C.A.; Aiola, S.; et al.
\newblock {Observation of a $\Lambda_b^0-\overline{\Lambda}_b^0$ production
 asymmetry in proton--proton collisions at $\sqrt{s} = 7 \textrm{ and }
 8\,\textrm{TeV}$}.
\newblock {\em JHEP} {\bf 2021}, {\em 10},~60.
 \href{http://xxx.lanl.gov/abs/2107.09593}
\newblock
 {\changeurlcolor{black}\href{https://doi.org/10.1007/JHEP10(2021)060}{\detokenize{https://doi.org/10.1007/JHEP10(2021)060}}}.

\bibitem[Aaij \em{et~al.}(2014)Aaij et~al.]{LHCb:2014kcb}
Aaij, R.; Adeva, B.; Adinolfi, M.; Affolder, A.; Ajaltouni, Z.; Albrecht, J.; Alessio, F.; Alexander, M.; Ali, S.; \linebreak Alkhazov, G.; et al.
\newblock {Measurement of $CP$ asymmetry in $D^0 \rightarrow K^- K^+$ and $D^0
 \rightarrow \pi^- \pi^+$ decays}.
\newblock {\em JHEP} {\bf 2014}, {\em 7},~41.
 \href{http://xxx.lanl.gov/abs/1405.2797}
\newblock
 {\changeurlcolor{black}\href{https://doi.org/10.1007/JHEP07(2014)041}{\detokenize{https://doi.org/10.1007/JHEP07(2014)041}}}.

\bibitem[Aaij \em{et~al.}(2012)Aaij et~al.]{LHCb:2012swq}
Aaij, R.; Beteta, C.A.; Adametz, A.; Adeva, B.; Adinolfi, M.; Adrover, C.; Affolder, A.; Ajaltouni, Z.; Albrecht, J.; \mbox{Alessio, F.; et al.}
\newblock {Measurement of the $D_s^+ - D_s^-$ production asymmetry in 7 TeV
 $pp$ collisions}.
\newblock {\em Phys. Lett. B} {\bf 2012}, {\em 713},~186--195.
 \href{http://xxx.lanl.gov/abs/1205.0897}
\newblock
 {\changeurlcolor{black}\href{https://doi.org/10.1016/j.physletb.2012.06.001}{\detokenize{https://doi.org/10.1016/j.physletb.2012.06.001}}}.

\bibitem[Bediaga and G\"obel(2020)]{Bediaga:2020qxg}
Bediaga, I.; G\"obel, C.
\newblock {Direct $CP$ violation in beauty and charm hadron decays}.
\newblock {\em Prog. Part. Nucl. Phys.} {\bf 2020}, {\em 114},~103808.
 \href{http://xxx.lanl.gov/abs/2009.07037}
\newblock
 {\changeurlcolor{black}\href{https://doi.org/10.1016/j.ppnp.2020.103808}{\detokenize{https://doi.org/10.1016/j.ppnp.2020.103808}}}.

\bibitem[Bediaga \em{et~al.}(2012)Bediaga, Miranda, dos Reis, Bigi, Gomes,
 Otalora~Goicochea, and Veiga]{Bediaga:2012tm}
Bediaga, I.; Miranda, J.; dos Reis, A.C.; Bigi, I.I.; Gomes, A.;
 Otalora~Goicochea, J.M.; Veiga, A.
\newblock {Second Generation of `Miranda Procedure' for CP Violation in Dalitz
 Studies of $B$ (and $D$ and $\tau$) Decays}.
\newblock {\em Phys. Rev. D} {\bf 2012}, {\em 86},~036005.
\newblock
 {\changeurlcolor{black}\href{https://doi.org/10.1103/PhysRevD.86.036005}{\detokenize{https://doi.org/10.1103/PhysRevD.86.036005}}}.

\bibitem[LHC(2022)]{LHCb:2022nyw}
Cheng, H.Y.; Chua, C.K.; Zhang, Z.Q. {Direct $CP$ violation in charmless three-body decays of $B^{\pm}$ mesons} \emph{Phys. Rev. D} \textbf{2016}, \mbox{\emph{94}, 094015.}


\bibitem[Zech and Aslan(2005)]{ET1}
Zech, G.; Aslan, B.
\newblock {A new test for the multivariate two-sample problem based on the
 concept of minimum energy}.
\newblock {\em J. Stat. Comput. Simul.} {\bf 2005}, {\em 75},~109.
 \href{http://xxx.lanl.gov/abs/0309164}
\newblock
 {\changeurlcolor{black}\href{https://doi.org/10.48550/arXiv.math/0309164}{\detokenize{https://doi.org/10.48550/arXiv.math/0309164}}}.

\bibitem[Henze(1998)]{kNN1}
Henze, N.
\newblock {A multivariate two-sample test based on the number of nearest
 neighbor type coincidences}.
\newblock {\em Ann. Stat.} {\bf 1998}, {\em 16},~772.

\bibitem[Williams(2010)]{Williams:2010vh}
Williams, M.
\newblock {How good are your fits? Unbinned multivariate goodness-of-fit tests
 in high energy physics}.
\newblock {\em JINST} {\bf 2010}, {\em 5},~P09004.
 \href{http://xxx.lanl.gov/abs/1006.3019}
\newblock
 {\changeurlcolor{black}\href{https://doi.org/10.1088/1748-0221/5/09/P09004}{\detokenize{https://doi.org/10.1088/1748-0221/5/09/P09004}}}.

\bibitem[Back \em{et~al.}(2018)Back et~al.]{Back:2017zqt}
Back, J.; Gershon, T.; Harrison, P.; Latham, T.; O’Hanlon, D.; Qian, W.; Sanchez, P.d.; Craik, D.; Ilic, J.; Goicochea, J.M.O.; et~al.
\newblock {LAURA$^{++}$: A Dalitz plot fitter}.
\newblock {\em Comput. Phys. Commun.} {\bf 2018}, {\em 231},~198--242.
 \href{http://xxx.lanl.gov/abs/1711.09854}
\newblock
 {\changeurlcolor{black}\href{https://doi.org/10.1016/j.cpc.2018.04.017}{\detokenize{https://doi.org/10.1016/j.cpc.2018.04.017}}}.

\bibitem[Jacob and Wick(1959)]{Jacob:1959at}
Jacob, M.; Wick, G.C.
\newblock {\textls[-10]{On the General Theory of Collisions for Particles with Spin}}.
\newblock {\em Ann. Phys.} {\bf 1959}, {\em 7},~404--428.
\newblock
 {\changeurlcolor{black}\href{https://doi.org/10.1016/0003-4916(59)90051-X}{\detokenize{https://doi.org/10.1016/0003-4916(59)90051-X}}}.

\bibitem[Zemach(1965)]{Zemach:1965ycj}
Zemach, C.
\newblock {Use of angular momentum tensors}.
\newblock {\em Phys. Rev.} {\bf 1965}, {\em 140},~B97--B108.
\newblock
 {\changeurlcolor{black}\href{https://doi.org/10.1103/PhysRev.140.B97}{\detokenize{https://doi.org/10.1103/PhysRev.140.B97}}}.

\bibitem[Aaij \em{et~al.}(2020)Aaij et~al.]{LHCb:2019sus}
Aaij, R.; Beteta, C.A.; Adametz, A.; Adeva, B.; Adinolfi, M.; Adrover, C.; Affolder, A.; Ajaltouni, Z.; Albrecht, J.; Alessio, F.; et al.
\newblock {Amplitude analysis of the $B^+ \rightarrow \pi^+\pi^+\pi^-$ decay}.
\newblock {\em Phys. Rev. D} {\bf 2020}, {\em 101},~012006.
 \href{http://xxx.lanl.gov/abs/1909.05212}
\newblock
 {\changeurlcolor{black}\href{https://doi.org/10.1103/PhysRevD.101.012006}{\detokenize{https://doi.org/10.1103/PhysRevD.101.012006}}}.

\bibitem[Zhang(2022)]{Zhang:2022emj}
Zhang, Z.H.
\newblock {Analysis of the angular distribution asymmetries and the associated
 $CP$ asymmetries in bottom baryon decays}. \emph{arXiv} {\bf 2022}, arXiv:2208.13411.
\newblock 

\bibitem[Bensalem \em{et~al.}(2002)Bensalem, Datta, and
 London]{TPA_Bensalem:2002ys}
Bensalem, W.; Datta, A.; London, D.
\newblock {New physics effects on triple product correlations in $\Lambda_b$
 decays}.
\newblock {\em Phys. Rev. D} {\bf 2002}, {\em 66},~094004.
 \href{http://xxx.lanl.gov/abs/hep-ph/0208054}
\newblock
 {\changeurlcolor{black}\href{https://doi.org/10.1103/PhysRevD.66.094004}{\detokenize{https://doi.org/10.1103/PhysRevD.66.094004}}}.

\bibitem[Gronau and Rosner(2015)]{TPA_Gronau:2015gha}
Gronau, M.; Rosner, J.L.
\newblock {Triple product asymmmetries in $\Lambda_b$ and $\Xi_b$ decays}.
\newblock {\em Phys. Lett. B} {\bf 2015}, {\em 749},~104--107.
 \href{http://xxx.lanl.gov/abs/1506.01346}
\newblock
 {\changeurlcolor{black}\href{https://doi.org/10.1016/j.physletb.2015.07.060}{\detokenize{https://doi.org/10.1016/j.physletb.2015.07.060}}}.

\bibitem[Geng and Liu(2021)]{Geng:2021sxe}
Geng, C.Q.; Liu, C.W.
\newblock {Time-reversal asymmetries and angular distributions in
 \ensuremath{\Lambda}$_{b}$ \textrightarrow{} \ensuremath{\Lambda}V}.
\newblock {\em JHEP} {\bf 2021}, {\em 11},~104.
 \href{http://xxx.lanl.gov/abs/2109.09524}
\newblock
 {\changeurlcolor{black}\href{https://doi.org/10.1007/JHEP11(2021)104}{\detokenize{https://doi.org/10.1007/JHEP11(2021)104}}}.

\bibitem[Wang \em{et~al.}(2022)Wang, Qin, and Yu]{Wang:2022fih}
Wang, J.P.; Qin, Q.; Yu, F.S.
\newblock {CP violation induced by T-odd correlations and its baryonic
 application}. \emph{arXiv} {\bf 2022}, arXiv:2211.07332.

\bibitem[Workman \em{et~al.}(2022)Workman et~al.]{PDG}
Particle Data Group; Workman, R.L.; Burkert, V.D.; Crede, V.; Klempt, E.; Thoma, U.; Tiator, L.; Agashe, K.; Aielli, G.; Allanach, B.C.; et al.
\newblock {Review of Particle Physics}.
\newblock {\em PTEP} {\bf 2022}, {\em 2022},~083C01.
\newblock
 {\changeurlcolor{black}\href{https://doi.org/10.1093/ptep/ptac097}{\detokenize{https://doi.org/10.1093/ptep/ptac097}}}.

\bibitem[Hsiao and Geng(2015)]{Hsiao:2014mua}
Hsiao, Y.K.; Geng, C.Q.
\newblock {Direct CP violation in $\Lambda_b$ decays}.
\newblock {\em Phys. Rev. D} {\bf 2015}, {\em 91},~116007.
 \href{http://xxx.lanl.gov/abs/1412.1899}
\newblock
\linebreak {\changeurlcolor{black}\href{https://doi.org/10.1103/PhysRevD.91.116007}{\detokenize{https://doi.org/10.1103/PhysRevD.91.116007}}}.

\bibitem[Lu \em{et~al.}(2009)Lu, Wang, Zou, Ali, and Kramer]{Lu:2009cm}
Lu, C.D.; Wang, Y.M.; Zou, H.; Ali, A.; Kramer, G.
\newblock {Anatomy of the pQCD Approach to the Baryonic Decays $\Lambda_b\to
 p\pi, pK$}.
\newblock {\em Phys. Rev. D} {\bf 2009}, {\em 80},~034011.
 \href{http://xxx.lanl.gov/abs/0906.1479}
\newblock
 {\changeurlcolor{black}\href{https://doi.org/10.1103/PhysRevD.80.034011}{\detokenize{https://doi.org/10.1103/PhysRevD.80.034011}}}.

\bibitem[Aaij \em{et~al.}(2018)Aaij et~al.]{LHCb-PAPER-2018-025}
Aaij, R.; Adeva, B.; Adinolfi, M.; Affolder, A.; Ajaltouni, Z.; Albrecht, J.; Alessio, F.; Alexander, M.; Ali, S.; \mbox{Alkhazov, G.; et al.}
\newblock {Search for $C\!P$ violation in $\Lambda^0_b \to p K^-$ and
 $\Lambda^0_b \to p \pi^-$ decays}.
\newblock {\em Phys. Lett. B} {\bf 2018}, {\em 787},~124--133.
\newblock
 \linebreak {\changeurlcolor{black}\href{https://doi.org/10.1016/j.physletb.2018.10.039}{\detokenize{https://doi.org/10.1016/j.physletb.2018.10.039}}}.

\bibitem[Aaij \em{et~al.}(2021{\natexlab{a}})Aaij et~al.]{LHCb-PAPER-2020-017}
Aaij, R.; Adeva, B.; Adinolfi, M.; Affolder, A.; Ajaltouni, Z.; Albrecht, J.; Alessio, F.; Alexander, M.; Ali, S.; Alkhazov, G.; et al.
\newblock {Search for $CP$ violation in $\Xi^-_b \to p K^- K^-$decays}.
\newblock {\em Phys. Rev. D} {\bf 2021}, {\em 104},~052010.
 {\changeurlcolor{black}\href{https://doi.org/10.1103/PhysRevD.104.052010}{\detokenize{https://doi.org/10.1103/PhysRevD.104.052010}}}.

\bibitem[Aaij \em{et~al.}(2021{\natexlab{b}})Aaij et~al.]{LbToD0pk_nonresonant}
Aaij, R.; Adeva, B.; Adinolfi, M.; Affolder, A.; Ajaltouni, Z.; Albrecht, J.; Alessio, F.; Alexander, M.; Ali, S.; Alkhazov, G.; et al.
\newblock {Observation of the suppressed $\Lambda_b^0\to Dp K^-$ decay with
 $D\to K^+\pi^-$ and measurement of its CP asymmetry}.
\newblock {\em Phys. Rev. D} {\bf 2021}, {\em 104},~112008.
 \href{http://xxx.lanl.gov/abs/2109.02621}
\newblock
 {\changeurlcolor{black}\href{https://doi.org/10.1103/PhysRevD.104.112008}{\detokenize{https://doi.org/10.1103/PhysRevD.104.112008}}}.

\bibitem[Zhang \em{et~al.}(2021)Zhang, Jiang, Chen, and Qian]{Zhang:2021sit}
Zhang, S.; Jiang, Y.; Chen, Z.; Qian, W.
\newblock {Sensitivity studies on the CKM angle $\gamma$ in $\Lambda_b^0 \to
 D\Lambda$ decays}. \emph{arXiv} {\bf 2021}, arXiv:2112.12954.


\bibitem[Aaij \em{et~al.}(2016)Aaij et~al.]{LHCb-PAPER-2016-004}
Aaij, R.; Adeva, B.; Adinolfi, M.; Affolder, A.; Ajaltouni, Z.; Albrecht, J.; Alessio, F.; Alexander, M.; Ali, S.; Alkhazov, G.; et al.
\newblock {Observations of $\Lambda_b^0 \to \Lambda K^+\pi^-$ and $\Lambda_b^0
 \to \Lambda K^+K^-$ decays and searches for other $\Lambda_b^0$ and $\Xi_b^0$
 decays to $\Lambda h^+h^{\prime -}$ final states}.
\newblock {\em JHEP} {\bf 2016}, {\em 5},~81,
 \href{http://xxx.lanl.gov/abs/1603.00413}
\newblock
 {\changeurlcolor{black}\href{https://doi.org/10.1007/JHEP05(2016)081}{\detokenize{https://doi.org/10.1007/JHEP05(2016)081}}}.

\bibitem[Aaij \em{et~al.}(2014)Aaij et~al.]{LHCb-PAPER-2013-061}
Aaij, R.; Adeva, B.; Adinolfi, M.; Affolder, A.; Ajaltouni, Z.; Albrecht, J.; Alessio, F.; Alexander, M.; Ali, S.; Alkhazov, G.; et al.
\newblock {Searches for $\Lambda^0_{b}$ and $\Xi^{0}_{b}$ decays to $K^0_{\rm
 S} p \pi^{-}$ and $K^0_{\rm S}p K^{-}$ final states with first observation of
 the $\Lambda^0_{b} \rightarrow K^0_{\rm S}p \pi^{-}$ decay}.
\newblock {\em JHEP} {\bf 2014}, {\em 4},~87.
 \href{http://xxx.lanl.gov/abs/1402.0770}
\newblock
 {\changeurlcolor{black}\href{https://doi.org/10.1007/JHEP04(2014)087}{\detokenize{https://doi.org/10.1007/JHEP04(2014)087}}}.

\bibitem[Aaij \em{et~al.}(2019)Aaij et~al.]{LHCb-PAPER-2018-044}
Aaij, R.; Adeva, B.; Adinolfi, M.; Affolder, A.; Ajaltouni, Z.; Albrecht, J.; Alessio, F.; Alexander, M.; Ali, S.; \mbox{Alkhazov, G.; et al.}
\newblock {Measurements of $CP$ asymmetries in charmless four-body
 $\Lambda_b^0$ and $\Xi_b^0$ decays}.
\newblock {\em Eur. Phys. J. C} {\bf 2019}, {\em 79},~745.
 \href{http://xxx.lanl.gov/abs/1903.06792}
\newblock
 \linebreak {\changeurlcolor{black}\href{https://doi.org/10.1140/epjc/s10052-019-7218-1}{\detokenize{https://doi.org/10.1140/epjc/s10052-019-7218-1}}}.

\bibitem[Aaij \em{et~al.}(2020)Aaij et~al.]{LHCb-PAPER-2019-028}
Aaij, R.; Adeva, B.; Adinolfi, M.; Affolder, A.; Ajaltouni, Z.; Albrecht, J.; Alessio, F.; Alexander, M.; Ali, S.; \mbox{Alkhazov, G.; et al.}
\newblock {Search for $CP$ violation and observation of $P$ violation in
 $\Lambda_b^0 \to p \pi^- \pi^+ \pi^-$ decays}.
\newblock {\em Phys. Rev. D} {\bf 2020}, {\em 102},~051101.
 \href{http://xxx.lanl.gov/abs/1912.10741}
\newblock
 {\changeurlcolor{black}\href{https://doi.org/10.1103/PhysRevD.102.051101}{\detokenize{https://doi.org/10.1103/PhysRevD.102.051101}}}.

\bibitem[Aaij \em{et~al.}(2022)Aaij et~al.]{LHCb-PAPER-2021-030}
Aaij, R.; Adeva, B.; Adinolfi, M.; Affolder, A.; Ajaltouni, Z.; Albrecht, J.; Alessio, F.; Alexander, M.; Ali, S.; \mbox{Alkhazov, G.; et al.}
\newblock {Measurement of the photon polarization in
 $\Lambda_{b}^{0}\to\Lambda\gamma$ decays}.
\newblock {\em Phys. Rev. D} {\bf 2022}, {\em 105},~L051104.
 \href{http://xxx.lanl.gov/abs/2111.10194}
\newblock
 \linebreak {\changeurlcolor{black}\href{https://doi.org/10.1103/PhysRevD.105.L051104}{\detokenize{https://doi.org/10.1103/PhysRevD.105.L051104}}}.

\bibitem[Chen \em{et~al.}(2021)Chen, Li, Qian, Shen, Xie, Yang, Zhang, and
 Zhang]{Chen:2021ftn}
Chen, S.; Li, Y.; Qian, W.; Shen, Z.; Xie, Y.; Yang, Z.; Zhang, L.; Zhang, Y.
\newblock {Heavy Flavour Physics and CP Violation at LHCb: A Ten-Year Review}.
 \emph{arXiv} {\bf 2021}, arXiv:2111.14360.


\bibitem[Aaij \em{et~al.}(2017)Aaij et~al.]{LHCB-PAPER-2016-059}
Aaij, R.; Adeva, B.; Adinolfi, M.; Affolder, A.; Ajaltouni, Z.; Albrecht, J.; Alessio, F.; Alexander, M.; Ali, S.; Alkhazov, G.; et al.
\newblock {Observation of the decay $\Lambda^0_b \to p K^- \mu^+ \mu^-$ and a
 search for $C\!P$ violation}.
\newblock {\em JHEP} {\bf 2017}, {\em 6},~108.
 \href{http://xxx.lanl.gov/abs/1703.00256}
\newblock
 {\changeurlcolor{black}\href{https://doi.org/10.1007/JHEP06(2017)108}{\detokenize{https://doi.org/10.1007/JHEP06(2017)108}}}.

\bibitem[Paracha \em{et~al.}(2015)Paracha, Ahmed, and Aslam]{Paracha:2014hca}
Paracha, M.A.; Ahmed, I.; Aslam, M.J.
\newblock {Imprints of CP violation asymmetries in rare $\Lambda_{b}\to
 \Lambda\ell^{+}\ell^{-}$ decay in family non-universal $Z^{\prime}$ model}.
\newblock {\em PTEP} {\bf 2015}, {\em 2015},~033B04.
 \href{http://xxx.lanl.gov/abs/1408.4318}
\newblock
 {\changeurlcolor{black}\href{https://doi.org/10.1093/ptep/ptv017}{\detokenize{https://doi.org/10.1093/ptep/ptv017}}}.

\bibitem[Alok \em{et~al.}(2011)Alok, Datta, Dighe, Duraisamy, Ghosh, and
 London]{Alok:2011gv}
Alok, A.K.; Datta, A.; Dighe, A.; Duraisamy, M.; Ghosh, D.; London, D.
\newblock {New Physics in $b\to s\mu^+\mu^-$: CP-Violating Observables}.
\newblock {\em JHEP} {\bf 2011}, {\em 11},~122.
 \href{http://xxx.lanl.gov/abs/1103.5344}
\newblock
 {\changeurlcolor{black}\href{https://doi.org/10.1007/JHEP11(2011)122}{\detokenize{https://doi.org/10.1007/JHEP11(2011)122}}}.

\bibitem[Bigi(2012)]{Bigi:2012_Charm_baryon_CPV_expectations}
Bigi, I.I.
\newblock {Probing CP Asymmetries in Charm Baryons Decays}. \emph{arXiv} {\bf 2012}, arXiv:1206.4554.


\bibitem[Aaij \em{et~al.}(2018)Aaij et~al.]{LHCb-PAPER-2017-044}
Aaij, R.; Adeva, B.; Adinolfi, M.; Affolder, A.; Ajaltouni, Z.; Albrecht, J.; Alessio, F.; Alexander, M.; Ali, S.; Alkhazov, G.; et al.
\newblock {A measurement of the $CP$ asymmetry difference in
 $\varLambda_{c}^{+} \to pK^{-}K^{+}$ and $p\pi^{-}\pi^{+}$ decays}.
\newblock {\em JHEP} {\bf 2018}, {\em 3},~182.
 \href{http://xxx.lanl.gov/abs/1712.07051}
\newblock
 {\changeurlcolor{black}\href{https://doi.org/10.1007/JHEP03(2018)182}{\detokenize{https://doi.org/10.1007/JHEP03(2018)182}}}.

\bibitem[Aaij \em{et~al.}(2020)Aaij et~al.]{LHCB-PAPER-2019-026}
Aaij, R.; Adeva, B.; Adinolfi, M.; Affolder, A.; Ajaltouni, Z.; Albrecht, J.; Alessio, F.; Alexander, M.; Ali, S.; \mbox{Alkhazov, G.; et al.}
\newblock {Search for $CP$ violation in ${{{\varXi }} ^+_{c}} \rightarrow {p}
 {{K} ^-} {{\pi } ^+} $ decays using model-independent techniques}.
\newblock {\em Eur. Phys. J. C} {\bf 2020}, {\em 80},~986.
 \href{http://xxx.lanl.gov/abs/2006.03145}
\newblock
 {\changeurlcolor{black}\href{https://doi.org/10.1140/epjc/s10052-020-8365-0}{\detokenize{https://doi.org/10.1140/epjc/s10052-020-8365-0}}}.

\end{thebibliography}
\end{document}